**Processes and characteristics of methane hydrate formation and decomposition: a microfluidic experimental study**

Yuze Wang[a*], Jianyu Yang[b], Pengfei Wang[c], Jinlong Zhu[d], Yongshun John Chen[e]

a. Associate Professor, Department of Ocean Science and Engineering, Southern University of Science and Technology, Shenzhen 518005, P.R. China

b. PhD Student, Department of Ocean Science and Engineering, Southern University of Science and Technology, Shenzhen 518005, P.R. China

c. Research Associate Professor, Shenzhen Key Laboratory of Natural Gas Hydrate & Institute of Major Scientific Facilities for New Materials & Academy for Advanced Interdisciplinary Studies, Southern University of Science and Technology, Shenzhen, 518055, P.R. China

d. Associate Professor, Department of Physics, Shenzhen Key Laboratory of Natural Gas Hydrate & Institute of Major Scientific Facilities for New Materials, Southern University of Science and Technology, Shenzhen 518005, PR. China

e. Chair Professor, Department of Ocean Science and Engineering, Southern University of Science and Technology, Shenzhen 518005, P.R. China

*Corresponding author: Yuze Wang; email address: wangyz@sustech.edu.cn

**Abstract**

The formation and decomposition of methane hydrates, particularly in porous media such as subsea sediments, have attracted significant research interest due to their implications for energy production, storage, and safety in deep-sea environments. This study explores the process and characteristics of methane hydrates formation and decomposition using microfluidic technology to mimic natural conditions. By incorporating methylene blue, we enhanced phase differentiation, identifying five hydrate types: block, vein, point, membrane, and shell. These forms were influenced by the presence and movement of free gas, which shaped their development. Block and vein hydrates mainly formed in water-filled pores, while point and membrane hydrates appeared as coatings related to gas migration. Shell hydrates formed after gas relocation, filling pores. During dissociation, the presence of free gas accelerated the process significantly, with a dissociation rate approximately 12 times faster than with water alone. Gas migration was key in accelerating hydrate breakdown and fragment formation. This research offers critical insights into methane hydrate behavior, aiding in optimizing natural gas extraction and preventing deep-sea pipeline blockages.

# 1. Introduction

The characteristics and kinetics of methane hydrate formation and decomposition have garnered significant research interest recently due to their potential for optimizing energy production, storage, and transportation. Methane hydrates, widely distributed in subsea sediments on continental margins and in permafrost regions, are considered a significant future energy resource, potentially offering energy self-sufficiency for coastal nations (Almenningen et al., 2018). Understanding the characteristics and kinetics of methane hydrate decomposition is crucial for optimizing excavation procedures and improving efficiency. Additionally, the formation of methane hydrates and their distribution in porous medium sediment reservoirs affect the dissociation process, making the formation process equally important for developing methane excavation procedures. Methane hydrates have a high energy density, with 1 m³ of methane hydrate capable of storing 180 m³ (STP) of methane, making them promising materials for natural gas storage and transportation (Chen et al., 2017). Therefore, understanding the formation characteristics and kinetics is essential for improving energy storage efficiency. Moreover, as oil and gas extraction advances into deep-sea environments, pipelines often create conditions conducive to hydrate formation, which can obstruct the pipelines and lead to severe accidents (Feng et al., 2024). Thus, understanding the formation and dissociation processes of methane hydrates is vital for finding ways to prevent their formation in pipelines.

Microfluidics, a technology for manipulating small fluid volumes within micrometer-scale channels (Whitesides, 2006), effectively mimics natural porous media and allows for precise control over critical experimental conditions such as temperature, pressure, and chemical composition. When combined with optical or confocal microscopy, microfluidics enhances research on biochemical reactions and multiphase reactive transport across fields such as analytical chemistry (Mitchell et al., 2001), biology (Kopp et al., 1998), materials science (Rattanarat et al., 2014; Alizadehgiashi et al., 2018), and various engineering disciplines (Wang et al., 2017; Wang et al., 2019). The application of microfluidics in gas hydrate research dates back to Tohidi et al. (2001). Microfluidics offers unique advantages for studying the growth, morphology, and dissociation characteristics of methane hydrates, as well as the multiphase flow related to hydrates. The spatial and temporal microscale distribution of gas, water, and hydrates can be observed, enabling the analysis of interactions during hydrate formation and dissociation.

Using microfluidics, it has been found that factors affecting hydrate formation and morphology include free gas and dissolved gas (Li et al., 2024; Zhang et al., 2024); the existing form of free methane bubbles (Li et al., 2022b); the gas-liquid interface (Ji et al., 2021); flow effects (Xu et al., 2023; Zhang et al., 2023; Rui et al., 2024); brine (Li et al., 2022b); resins (Feng et al., 2024); wettability of the porous medium (Ouyang et al., 2023); and the hydrate memory effect (Li et al., 2022a). Methane hydrate formation is generally divided into two types: hydrates formed from free gas at the gas-liquid interface and hydrates formed from dissolved gas (Tohidi et al., 2001). The first type of methane hydrate formation occurs at the gas-liquid interface and towards the gas phase (Ji et al., 2021; Zhang et al., 2023), and is sometimes considered to be formed from free gas (Zhang et al., 2023). The characteristics of these hydrates vary between studies. Li et al. (2022b) found that the hydrates depend on the size of the methane bubbles. Large methane bubbles trapped in pores make complete consumption difficult, leading to the formation of hydrate shells, whereas small methane bubbles can be entirely consumed for hydrate growth, resulting in the formation of transparent hydrate crystals. Other studies have found that free gas forms porous structure hydrates composed of many tiny methane hydrate crystals (Li et al., 2022b) and porous-type methane hydrates with rough surfaces (Zhang et al., 2024). The second type of methane hydrate, formed from dissolved gas, preferentially occurs on the surface of the hydrate crust formed from free gas and gradually grows toward the water phase (Li et al., 2024). These hydrates are smooth and transparent polyhedral hydrate crystals (Li et al., 2024) and crystal-type hydrates (Zhang et al., 2024).

Factors affecting hydrate dissociation and gas generation include the forms of hydrate formation (Li et al., 2022b) and gas-water migration (Yang et al., 2024). Shell hydrates, which enclose unconsumed methane bubbles, start to dissociate first, whereas transparent small hydrate crystals require a longer time and higher driving force for dissociation (Li et al., 2022b). The dissociation process can be categorized into three stages: (a) single gas bubble growth with an expanding water layer at an initial slow dissociation rate, (b) rapid generation of clusters of gas bubbles at a fast dissociation rate, and (c) gas bubble coalescence with uniform distribution in the pore space (Zhang et al., 2024). The formation and expansion of gas microbubbles under depressurization, along with gas-water migration, accelerate hydrate dissociation (Yang et al., 2024). Notably, depressurization-induced gas slug flow increases the dissociation rate by more than an order of magnitude compared to stationary gas-water conditions (Yang et al., 2024). In addition, at the initial stage of dissociation, hydrate blocks not only transform into water and gas but also decompose into small pieces. Gas-liquid interfaces first appear around the hydrate block edges, and shrinkage cavities form on the hydrate block surface during dissociation (Chen et al., 2019).

Although many scholars have researched the formation and decomposition processes of natural gas hydrates using microfluidic technology, their findings have sometimes been inconsistent. For example, the descriptions of methane hydrate formed from the gas phase vary between studies. Additionally, gas phases significantly affect both the formation and dissociation processes of methane gas hydrate, yet the quantified correlation between gas phase and gas hydrate formation and dissociation kinetics is not fully understood. This inconsistency and difficulty in quantifying different phases during methane hydrate formation and dissociation may be due to the similar appearances of water, gas, hydrate, and the microfluidic matrix, making precise observation challenging. In this study, following Tohidi et al. (2001), we add methylene blue to the liquid phase to easily distinguish the water phase from other phases and differentiate gas phases from hydrate phases based on shape changes over time. This method allows us to quantify changes in natural gas hydrate saturation, type, nucleation and growth processes, morphology, dissociation procedures, and the interactions of both gas and water with hydrates during formation and dissociation.

# 2. Material and Methods

### Microfluidic chip and experimental setup

A microfluidic chip with realistic sand grain characteristics was designed following Wang et al. (2019) and fabricated by photolithography and wet etching process on glass slide and low-temperature bonding process. The pore shape and size were designed based on cross-sectional image resembling pore network in quartz sandstone (Wang et al., 2019). Reproduction of actual pore bodies, pore necks, and coordination numbers made the model suitable for flow and equilibria studies related to natural sediments (Wang et al., 2019). The grainsize dimensions were 40 mm to 700 mm and average pore diameter of the porous medium was approximately 270 μm. The constant vertical height of microfluidic channels was 30 μm. The porosity of the microfluidic chip was about 40%. The wettability of the solid grains was strongly water-wet because of the anodic bonding procedure (Almenningen et al., 2018).

The schematic diagram illustrating the connections between each component is shown in Figure 1. The microfluidic chip, equipped with an inlet and outlet, is placed inside a high-pressure container that also has an inlet and outlet. The high-pressure container filled with water serves two functions. First, it maintains the pressure outside the microfluidic chip using an automatic confining pressure control pump connected to the inlet of the high-pressure container, which can be set to a value always higher than the pressure inside the microfluidic chip. This is necessary because the microfluidic chip without confinement cannot withstand the high pressure required for methane hydrate formation. However, with the assistance of external pressure, the pressure inside the microfluidic chip can be increased to form methane hydrate. The microfluidic chip can hold a compressive pressure of up to 2 MPa from the outside water. Therefore, the water pressure of the container is always set to be 2 MPa higher than the pressure inside the microfluidic chip. Second, it maintains the temperature of the microfluidic chip by regulating the water temperature outside the chip. This is achieved through a water bath system that circulates temperature-controlled water around the container throughout the entire experiment (water bath system not shown in Figure 1). The automatic confining pressure control pump, model TC-100D, is designed for a pressure capacity of 40 MPa, with a resolution of 0.001 mL and a flow rate control range of 0.001-30 mL/min. The inlet of the microfluidic chip is connected to a liquid injection pump and a gas injection pump to inject water and methane into the microfluidic chip, respectively. The liquid injection pump model is 2PB1040 by Ossich, with a maximum pressure capacity of 40 MPa and a flow rate control range of 0.01-9.9 mL/min. The gas injection pump model is TC-100D, designed for a pressure capacity of 70 MPa, with a resolution of 0.001 mL and a flow rate control range of 0.001-30 mL/min. The outlet of the microfluidic chip is connected to a back pressure control system. This system can set a pressure that, whenever the pressure is higher than the set value, the valve automatically opens to release the liquid inside the microfluidic chip. An optical microscope, model MZL-4KZD, is placed above the high-pressure container, and a lighting system is placed beneath it. Because both the top and bottom of the high-pressure container and the microfluidic chip are made of transparent material, light can pass through the container and the microfluidic chip. With the microscope, the methane hydrate formation and dissociation processes can be captured.

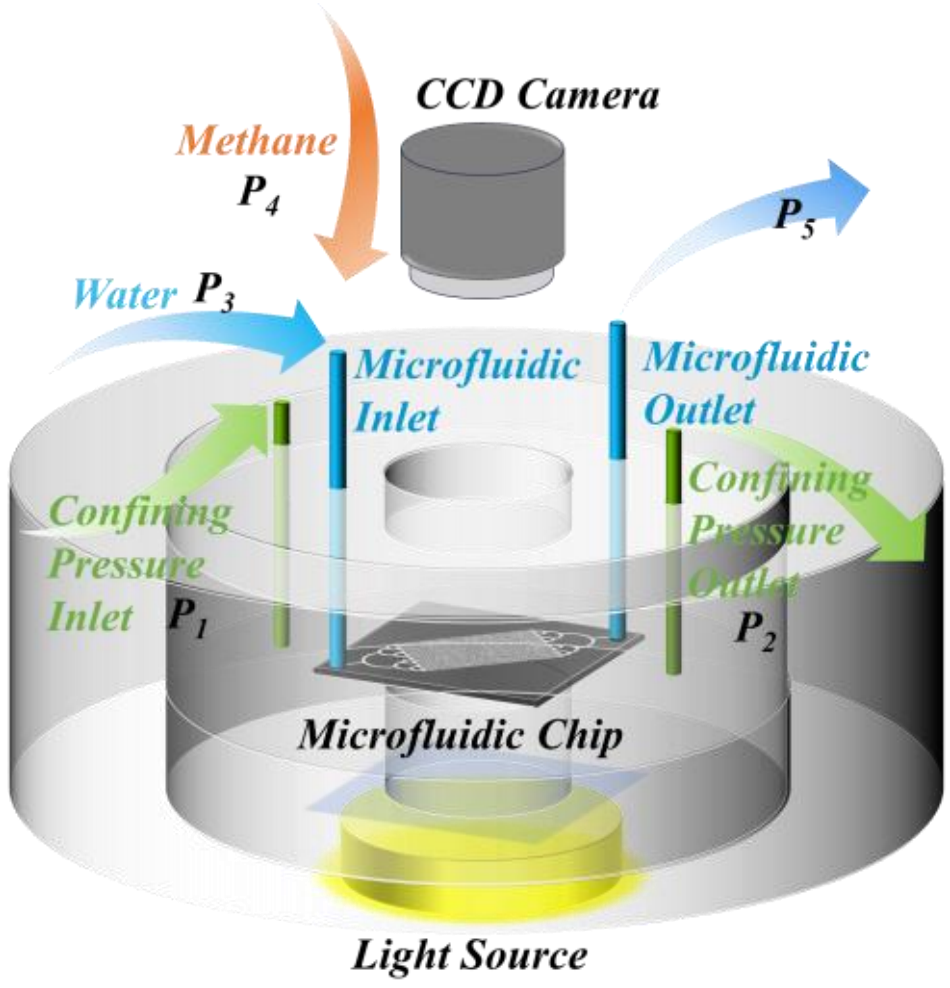

Figure 1 Schemic of microfluidic experiment setup for gas hydrate formation and dissociation

**Formation and decomposition of methane hydrates**

The experimental materials used in this study mainly include deionized water, methylene blue, and 99.9% pure methane (Shenzhen Huapeng Gas Co., Ltd.). 0.7% methylene blue is added to deionized water to make deionized water blue color under microscope.

To form methane hydrate inside the microfluidic chip, the following steps were undertaken: (1) The microfluidic chip was placed inside a high-pressure container, which was then fully sealed. The automatic confining pressure control pump was set to 0.3 MPa (P1 equals to 0.3 MPa in Figure 1), and the valve at the outlet of the high-pressure container was opened (P2 equals to 0. 1 MPa in Figure 1)to allow water to be injected. Once water was visible at the outlet, the valve was closed. (2) The pressure of the automatic confining pressure control pump was gradually increased from 0.3 MPa to 2.0 MPa, allowing water to enter the system and increasing the water pressure outside the microfluidic chip to 2.0 MPa. (3) A liquid injection pump connected to a solution of 0.7% methylene blue in deionized water was set to 0.5 ml/min. This solution was injected into the microfluidic chip through the microfluidic chip inlet (P3 in Figure 1). The outlet valve of the microfluidic chip was opened, allowing water to fill the chip. Once water was visible at the outlet, it indicated the chip was saturated with water. (4) The pressure (P5 in Figure 1) of the back-pressure control system was gradually increased to 8 MPa while continuously injecting water into the microfluidic chip until the internal pressure reached 8 MPa. Throughout this process, the pressure outside the microfluidic chip was maintained at 2 MPa higher than the internal pressure by injecting water into the high-pressure container. (5) The liquid injection pump was then stopped, and the gas injection pump was opened, with the gas pressure (P4 in Figure 1) set to 9 MPa. Gas was injected into the microfluidic chip for about 5 seconds and then stopped. The liquid injection pump was reopened to inject water into the microfluidic chip, facilitating the gas injection into the chip. Once the gas saturation in the microfluidic chip reaches the desined value, the water injection was stopped, indicating the coexistence of water and gas inside the chip. (6) To form gas hydrate, the temperature of the water bath was decreased from 20°C to 2°C over approximately 2 hours. During this temperature reduction, gas hydrates began to form.

Proposed gas production methods, including depressurization, thermal stimulation, chemical inhibitors, or $CO_2$ replacement, typically target methane gas hydrates, with depressurization being preferred due to its avoidance of fluid injection (Almenningen et al., 2018). Therefore, methane hydrate dissociation procedure during depression was selected in this study. During the depressurization process, the pressure of the back pressure control system was reduced from approximately 8 MPa to 3 MPa in about 2 minutes. After an additional 4 minutes, the pressure further decreased to around 2 MPa within 20 seconds.

**Imaging and phase distinguish**

The microscopic image of a microfluidic chip containing a porous medium matrix, free gas-phase methane, deionized water dyed with methylene blue, and methane hydrate is shown in Figure 2. The microfluidic porous medium matrix (M1-M7) remains unchanged during the formation and dissociation of gas hydrates, allowing for easy distinction. Although the gas and the microfluidic porous medium matrix have similar colors Figure 2b, the gas has a black border and changes in shape, size, and location during the formation and dissociation processes, making it distinguishable. The water phase is dyed blue, facilitating its identification. Methane hydrate (H1-H10 in Figure 2a) exists in different morphologies, changes during the formation and dissociation processes, and lacks the black border characteristic of gas, allowing it to be easily distinguished.

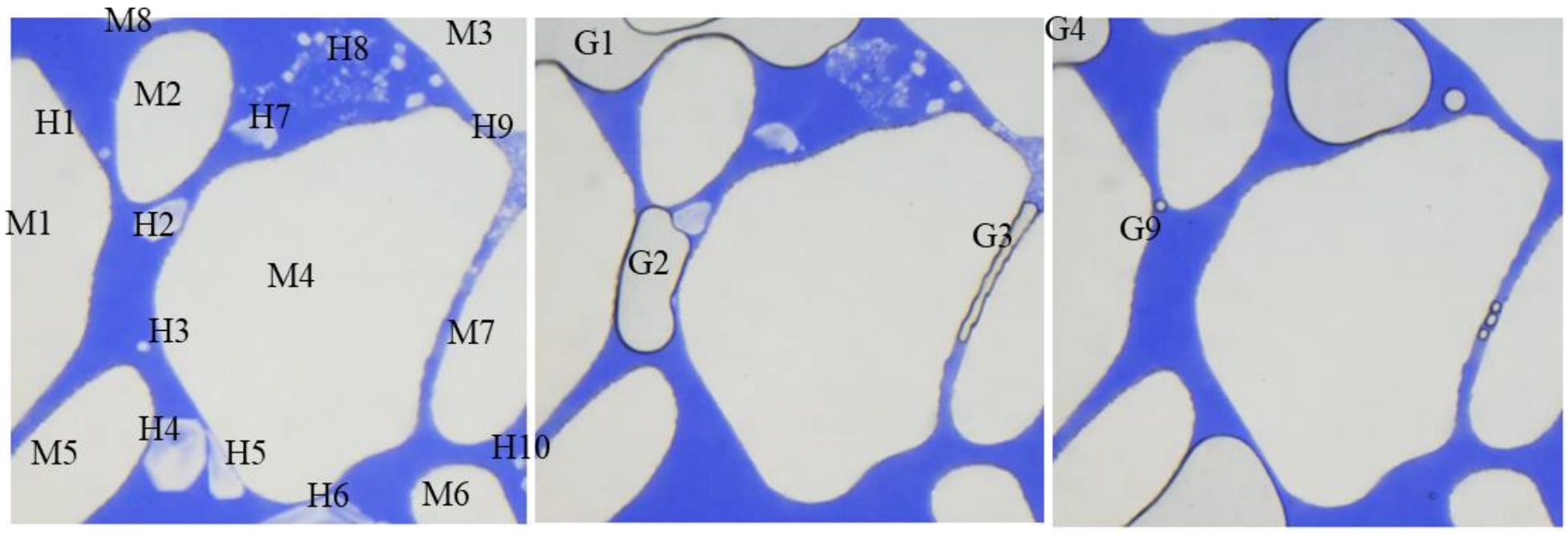

Figure 2 Microscopic images containing microfluidic porous medium matrix M1-M7, methane gas G1-G10, gas hydrate H1-H10, and water before (a), during (b) and after (c) hydrate dissociation

### Quantification of hydrate, gas and water

The microfluidics chip experiment offers the advantage of observing the entire formation and dissociation process of methane hydrate at the pore scale. However, quantifying the different phases has been challenging. In this study, several types of methane hydrates were observed, and their quantification varied based on morphology. For block hydrates, free gas phases, and water phases, the volume was quantified based on the area, assuming the hydrate occupies the entire depth of the microfluidic chip. For dot hydrates, assumed to be hemispherical, the volume was calculated using the diameter and the volume equation. For shell hydrates, the volume was quantified assuming the hydrate occupies the entire depth of the microfluidic chip. Additionally, theoretical values of methane hydrate, based on changes in free gas phases during the formation and dissociation of methane hydrate, were calculated and compared with the measured and quantified methane hydrate volume. The differences were discussed considering the occupying depths of the methane hydrate in the microfluidic chip and the contribution of dissolved methane in water.

# 3. Results and Discussion

### Morphology of methane hydrates

After injecting both methane gas and water into the microfluidic chip (Figure 3a), the free gas-phase and water-phase regions each accounted for about 40% and 60% (as shown in Figure 3a). Despite the size of the gas phases, the color in the images containing gas appears pure white, even if the gas does not fully occupy the entire depth of the microfluidic chip. Following the formation of methane hydrates, the free gas phase was completely consumed, resulting in the formation of five types of hydrates (Figures 3b and 3c): block hydrates, vein hydrates, point hydrates, shell hydrates, and film hydrates. Block hydrates (circled in black in Figures 3b and 3c) are primarily white with occasional light blue coloration and varying shapes. Vein hydrates (circled in green in Figures 3b and 3c), similar to block hydrates in color, but are large and grow through several pores, with their shapes influenced by the porous structure. Point hydrates (circled in yellow in Figures 3b and 3c) are small, white crystals sparsely distributed in one or several adjacent pores. Shell hydrates are crystals (circled in red in Figures 3b and 3c) with a white border and light blue interior. Membrane hydrates (circled in blue in Figures 3b and 3c), light blue in color, occupy a single pore. The color of the hydrate, unlike the gas phase, is either pure white or mixture of blue and white. This suggests that, depending on the thickness of the hydrate, water present in the same z-direction as the hydrate, which is the observation direction, can be seen through the hydrate.

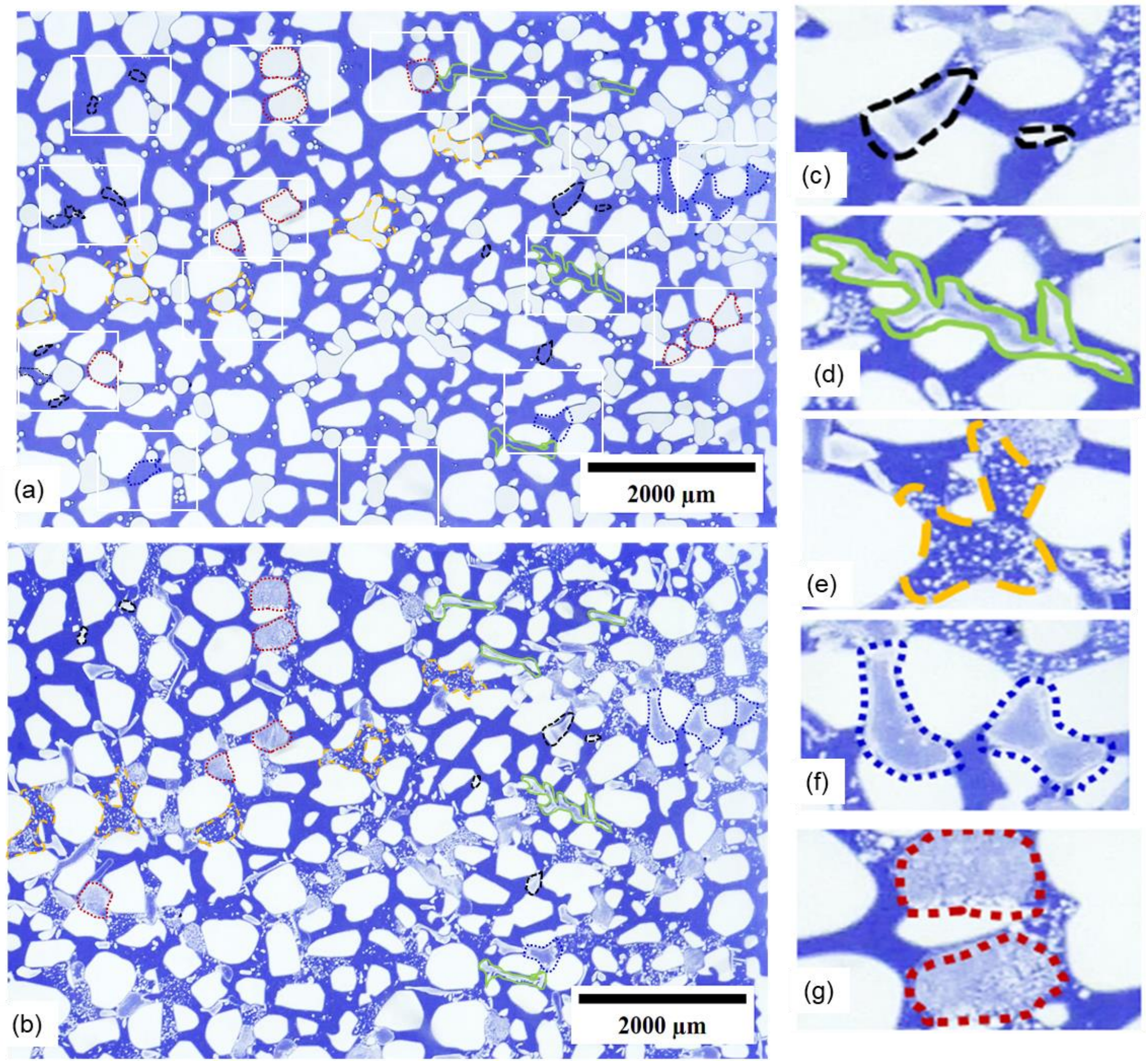

Figure 3 Microscopic images: (a) Before hydrate formation; (b) After hydrate formation; (c) Block hydrates; (d) Vein hydrates; (e) Point hydrates; (f) Shell hydrates; and (g) Film hydrates.

To observe the different hydrates in more detail, five areas of the images containing these hydrate types are shown in Figure 4. To compare the images before and after hydrate formation, images of the same area are also shown in Figure 4. Block and vein hydrates formed in areas with water phases, while point hydrates formed in areas with free gas phases, although the area size does not completely match the original gas phase area. Membrane hydrates also formed in gas phase regions but, compared to point hydrates, their shape more closely resembled the shape of the free gas phase. Shell hydrates formed where there were no free gas hydrates, similar to block and vein hydrates, but with point hydrates were always formed nearby. These observations indicate that the formation of different hydrate types is influenced by the pore structure, free gas phase, water phases, and the interactions between different hydrate phases.

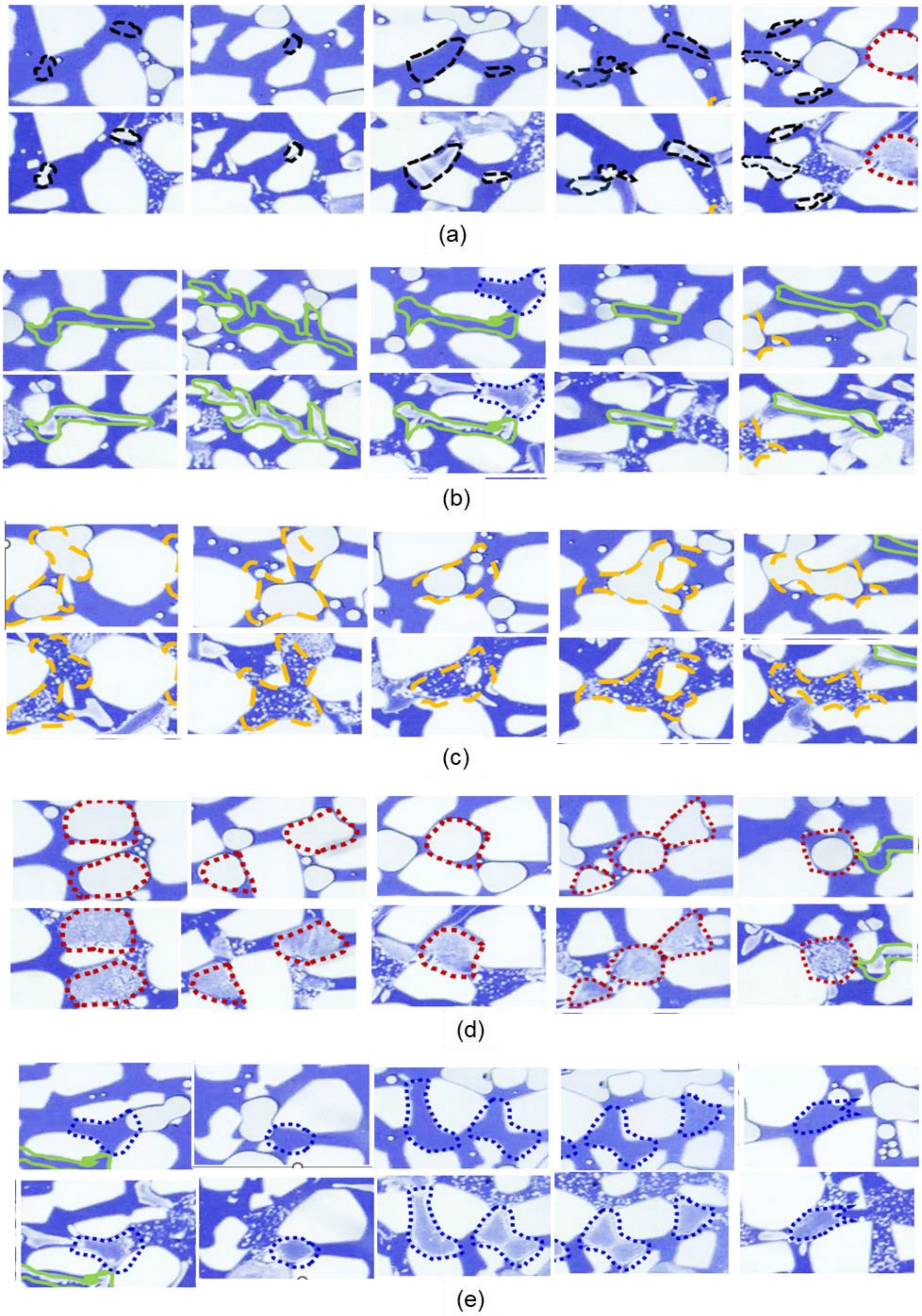

Figure 4 Microscopic images of a microfluidic chip taken at five areas, showing where (a) block hydrates, (b) vein hydrates, (c) point hydrates, (d) shell hydrates, and (e) film hydrates can form. The first row of images shows the areas before hydrate formation, and the second row shows the areas after hydrate formation.

Previous studies have reported the formation of rough hydrates from free gas and massive hydrates from dissolved gas (Ji et al., 2021; Li et al., 2022b; Zhang et al., 2023; Li et al., 2024; Zhang et al., 2024). Building on these studies, the following sections will provide a detailed analysis of the formation mechanisms of all five characteristic hydrate types observed in our current study, and how they are influenced by gas, liquid, and solid phases. This includes an examination of the formation processes of various hydrates, as well as the distribution and variation characteristics of the gas and liquid phases.

**Formation of block and vein hydrates from solid phase towards water phase**

The procedure of gas hydrate formation, similar to the crystallization of other materials, are generally described in two stages: nucleation followed by crystal growth (Chen et al., 2017). The formation of block and vein hydrates requires nucleation from a solid surface, which could be the inner surface of the microfluidic chip (Figure 5a) or pre-existing hydrates (Figure 5b). During the growth of these hydrates, there is no contact with the gas phase, only with the liquid phase (Figure 5). Therefore, it can be inferred that the methane for hydrate growth originates from the dissolved gas in the water. This observation is consistent with the findings of Tohidi et al. (2001), Hauge et al. (2016), Zhang et al. (2024) and Li et al. (2024), which indicate that hydrate initiation and growth can occur from gas dissolved in the water phase. Hydrates growing perpendicular to the observation direction (defined as the x-y plane) appear light blue (Figure 5 c-e), while those growing parallel to the observation direction (defined here as the z direction) appear white (Figure 5 c-e). Some crystals exhibit both growth directions simultaneously, resulting in uneven coloration within the same hydrate crystal (Figure 5 c-e). A comparison of growth times shows that hydrates in the x-y plane stop growing after 32 minutes (Figure 5c, $T_0$+1 min ~ $T_0$+33 min), whereas those in the z direction continue to grow for $T_0$+1594 minutes (Figure 5c, $T_0$+33 min ~ $T_0$+1627 min). This suggests that when free gas is present, it continuously dissolves methane into the water, particularly increasing the methane concentration in the x-y plane compared to the z direction, thereby promoting hydrate growth in the x-y plane. Consequently, hydrates growing in the x-y plane appear light blue because they are very thin, allowing the blue color of the water beneath them to show through. Once the free gas phase disappears, the methane concentration in the x-y plane decreases, and hydrate growth shifts to the z direction, resulting in white block hydrates, as the thicker hydrates obscure the underlying water color.

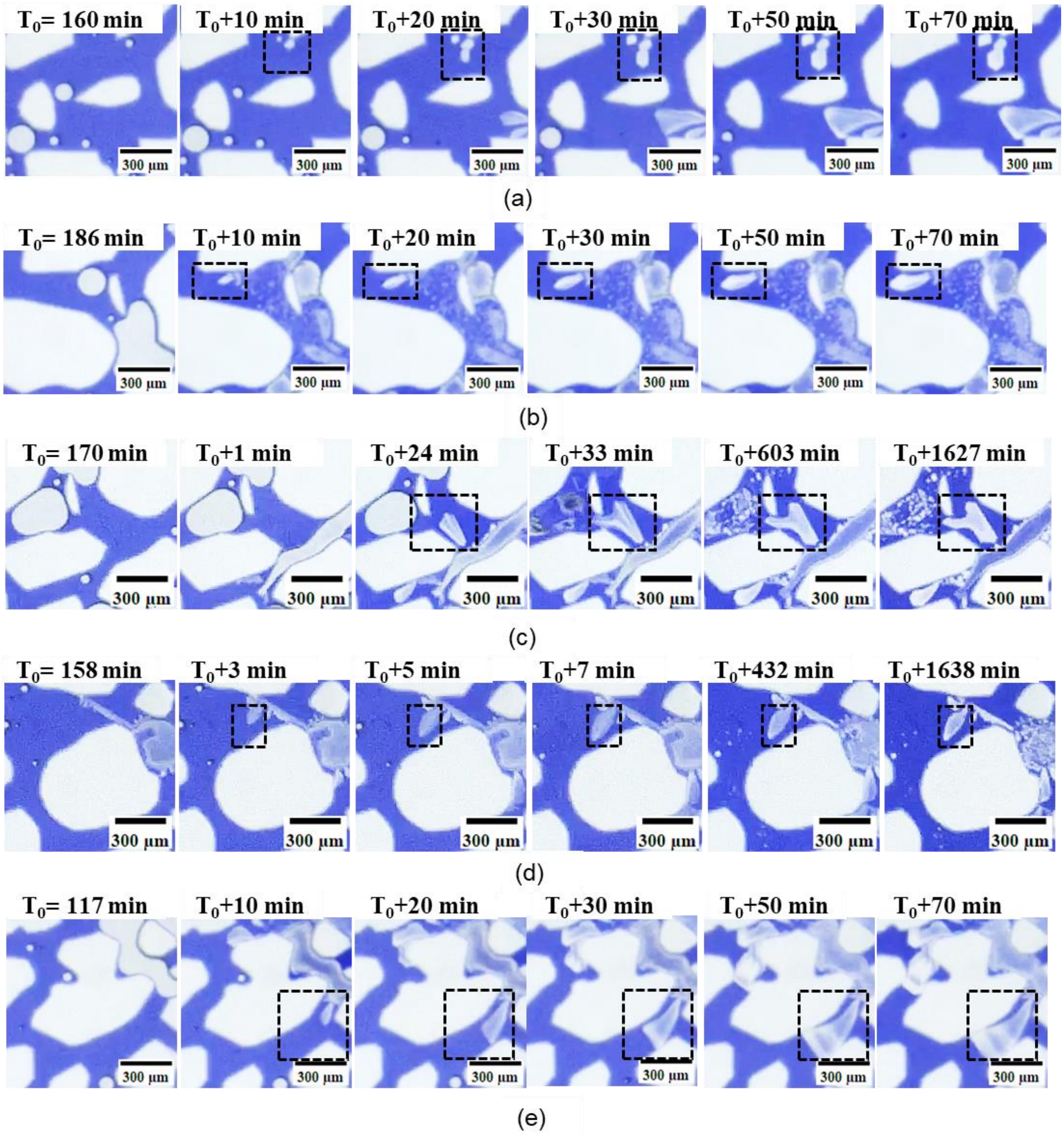

Figure 5 Nucleation and growth of block hydrate: (a) Nucleation in the microfluidic chip structure; (b) Nucleation on the hydrate surface; (c) Longitudinal growth of hydrate; (d) Lateral growth of hydrate; (e) Lateral and longitudinal growth of a single hydrate.

Block hydrates can continue to grow in the presence of sufficient free gas, eventually forming elongated vein or branch-like hydrates (Figure 6). The growth direction of vein hydrates is influenced by the location of the gas phase and the characteristics of the pore structure (Figure 6a, $T_0$+10 min G1, B1; 20 min G2, B2). If microfluidic porous structure pillars obstruct the growth direction, the hydrate will grow around them and continue toward the gas phase (Figure 6a, $T_0$+20 ~ $T_0$+30 min B2-B3, B4, B5). Upon reaching the gas phase, if the gas has not fully decomposed, the hydrate can induce gas migration, leading to the formation of new types of hydrates (more details can be found in sections on point and shell hydrates) (Figure 6b, $T_0$+150 ~ $T_0$+210 min H3). Hydrate growth is also influenced by surrounding hydrates and the locations of existing gas phases (Figure 6b, $T_0$ ~ $T_0$+30 min G1,

G2, G3), which together impact the growth direction and shape of the hydrates (Figure 6b, $T_0$+60 min H1, H2). When the direction of the gas phase aligns linearly with the hydrate growth direction, and there is no particle obstruction along the path (Figure 6b, $T_0$+60 ~ $T_0$+210 min G2, G3, H3), linear vein-like hydrates form (Figure 6b, $T_0$+60 ~ $T_0$+210 min H3). Even if other gas phases are present in the system (Figure 6b, $T_0$+90 min G1), if other hydrates lie between the gas phase and the existing hydrate (Figure 6b, $T_0$+ 90 min H1), the gas phases will preferentially support the growth of those hydrates (Figure 6b, $T_0$+90 ~$T_0$+150 min H1), thereby not affecting the growth direction and shape of the current hydrate (Figure 6b, $T_0$+90 ~ $T_0$+210 min H2).

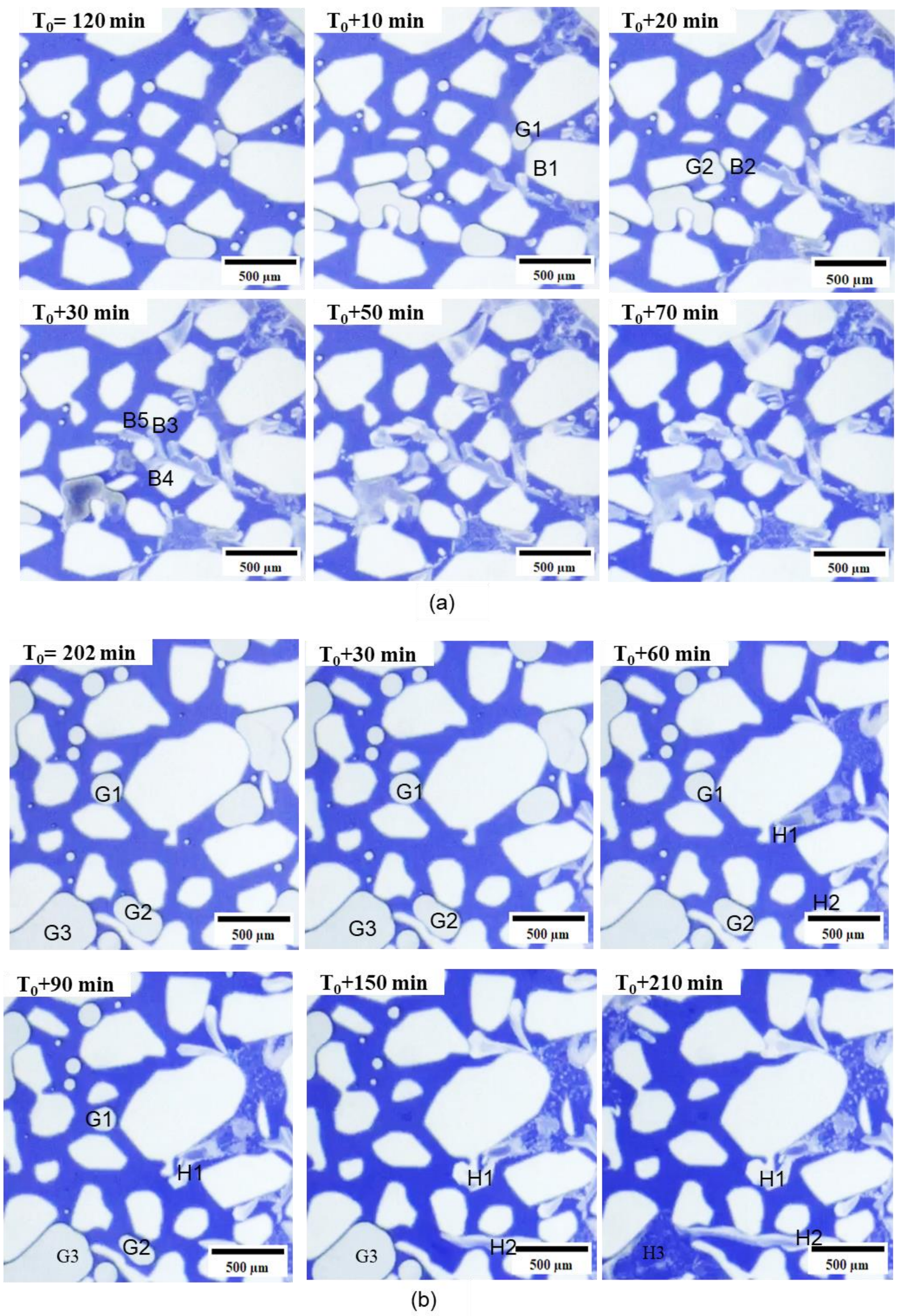

Figure 6. Formation process of (a) branch-like vein hydrate and (b) vein-like hydrate

Previous studies have found that hydrates grow from dissolved gas, initially forming from solid phases, such as the surface of a hydrate crust formed from free gas (Li et al., 2024). These hydrates initially grow away from the grain surface (Tohidi et al., 2001), moving toward the water phase (Li et al., 2024) and the center of the pores (Li et al., 2022a). As crystal growth continues, hydrates can gradually extend to the grain surface and eventually attach to it (Kuang et al., 2020), resulting in a combined grain-coating (Wang et al., 2021a), pore-filling (Wang et al., 2021a; Li et al., 2024), or load-bearing morphology in the sediments (Li et al., 2024). The hydrates formed are smooth, transparent polyhedral crystals (Li et al., 2024).

**Effects of free gas phases on hydrate formed from dissolved gas phases**

The formation of blocky or vein-like hydrates grow through liquid phase towards the free gas phase of methane (Figure 7a) indicates that the hydrate not only consume the dissolved methane but also the nearby gas-phase methane. This suggests that the growth of hydrates depletes the dissolved gas in the free water and as the concentration of dissolved gas in water phase decreases because of hydrate formation, the nearby free gas continuously dissolves into the free water, reducing the volume of the gas phase and sustaining the growth of hydrates. As the gas volume decreases (Figure 7a $T_0$ ~ $T_0$+40 min, G2-G4), the volume of hydrates increases (Figure 7a 0-40 min, H1-H4). After the gas phase disappears (Figure 7a $T_0$+60 min ~ $T_0$+100 min, G3-G4), hydrate growth ceases (Figure 7a $T_0$+40 min, H3-H4). The free gas phase (Figure7a $T_0$+60 min ~ $T_0$+140 min, G1-G2) preferentially supports the growth of adjacent hydrates (Figure7a $T_0$+60 min ~ $T_0$+140 min, H1-H2 compared to H3-H4). The areas of hydrate crystals and free gas in Figure 7a were quantified, with results presented in Figure 7b-e. As the area of the free gas phase decreased, the area occupied by the hydrates increased (Figure 7b). When the gas phase became smaller than the hydrate phase, the conversion efficiency of methane gas into block hydrates significantly decreased (Figure 7b and c). With the reduction in the gas phase area, the growth rate of hydrates exhibited distinct phases: steady high-speed growth, exponential growth, and slow growth (Figure 7d). The growth rate decreased linearly over time (Figure 7e). The measured growth rate of hydrates, based on the hydrate size, is generally lower than the calculated growth rate, which is based on the size change of the gas (Figure 7e). This discrepancy may be due to gas contributing to the formation of other hydrates or being dissolved in the water.

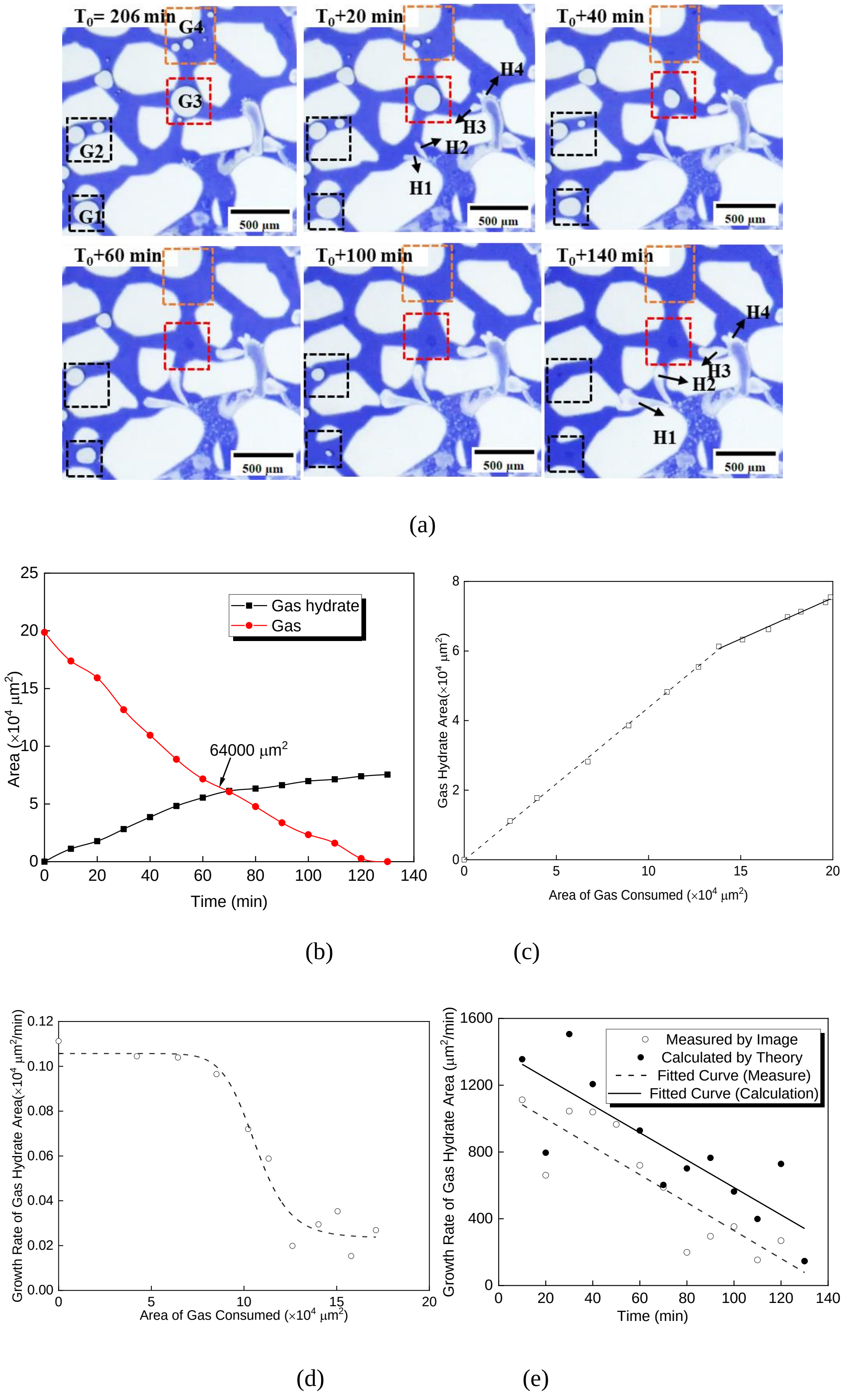

Figure 7. Free phase gas and block gas hydrate changes during hydrate formation (a) and quantification of block hydrate (b-e): (b) Changes in the area of block hydrate and gas phase

over time; (c) Relationship between block hydrate growth efficiency and gas consumption; (d) Relationship between hydrate growth rate and gas phase consumption; (e) Changes in hydrate formation rate over time.

This is consistent with the observations of crystalline growth where all the free methane gas was consumed and the pore was filled with solid nonporous methane gas hydrate (Almenningen et al., 2018), and the hydrate formation rate was determined by the concentration of methane molecules in surroundings water (Li et al., 2024). Building on this, this paper provides a detailed quantification of various parameters involved in the process, such as the change in hydrate area over time and the variation of the free gas phase over time.

**Formation of point hydrate in gas-water membrane at movable gas phase**

When block hydrates grow into the free gas phase, they induce the quick formation of hydrates within the water film between the gas phase and the microfluidic chip surface, where the water remains in prolonged contact with the free gas phase methane and contains a relatively high concentration of dissolved methane (Figure 8a, $T_0$ ~ $T_0$+0.29 s). Since hydrate growth requires both methane and water, and the water content in the film is limited, once the water is consumed, surrounding water invades the gas phase (Figure 8a, $T_0$+0.29- $T_0$+0.38 s), causing the unreacted gas to migrate to adjacent pores (Figure 8a, $T_0$+0.38- $T_0$+1.92 s). At the original gas phase location, a hydrate film forms (Figure 8a, $T_0$+1.92- $T_0$+303 s), which gradually develops into point hydrates over time (Figure 8a, 303 s-1618 min). Figure 8b and 8c shows the same hydrate formation and gas migration procedures. The formation of point hydrates, rather than block hydrates, occurs because although the water film becomes saturated with methane, its overall volume is small, and the amount of saturated methane is insufficient to form a continuous hydrate, resulting in the formation of point hydrates. In most of the previous studies, the point hydrates have been described as porous hydrates. For example, Rui et al. (2024) define this type of crystal porous and found that the faster the hydrate grows, the more porous it becomes. In addition, when hydrate blossomy growth occurs, the water phase could easily pass through the area and therefore, it was described that the porous hydrate has a greater porosity. Zhang et al. (2024) also described this type of hydrate as porous-type methane hydrate with a rough surface formed from methane gas bubbles at the gas–liquid interface. Following these studies, with the help of methylene blue, it is more clearly shown that the porous hydrates are individual point hydrates, with water surround them.

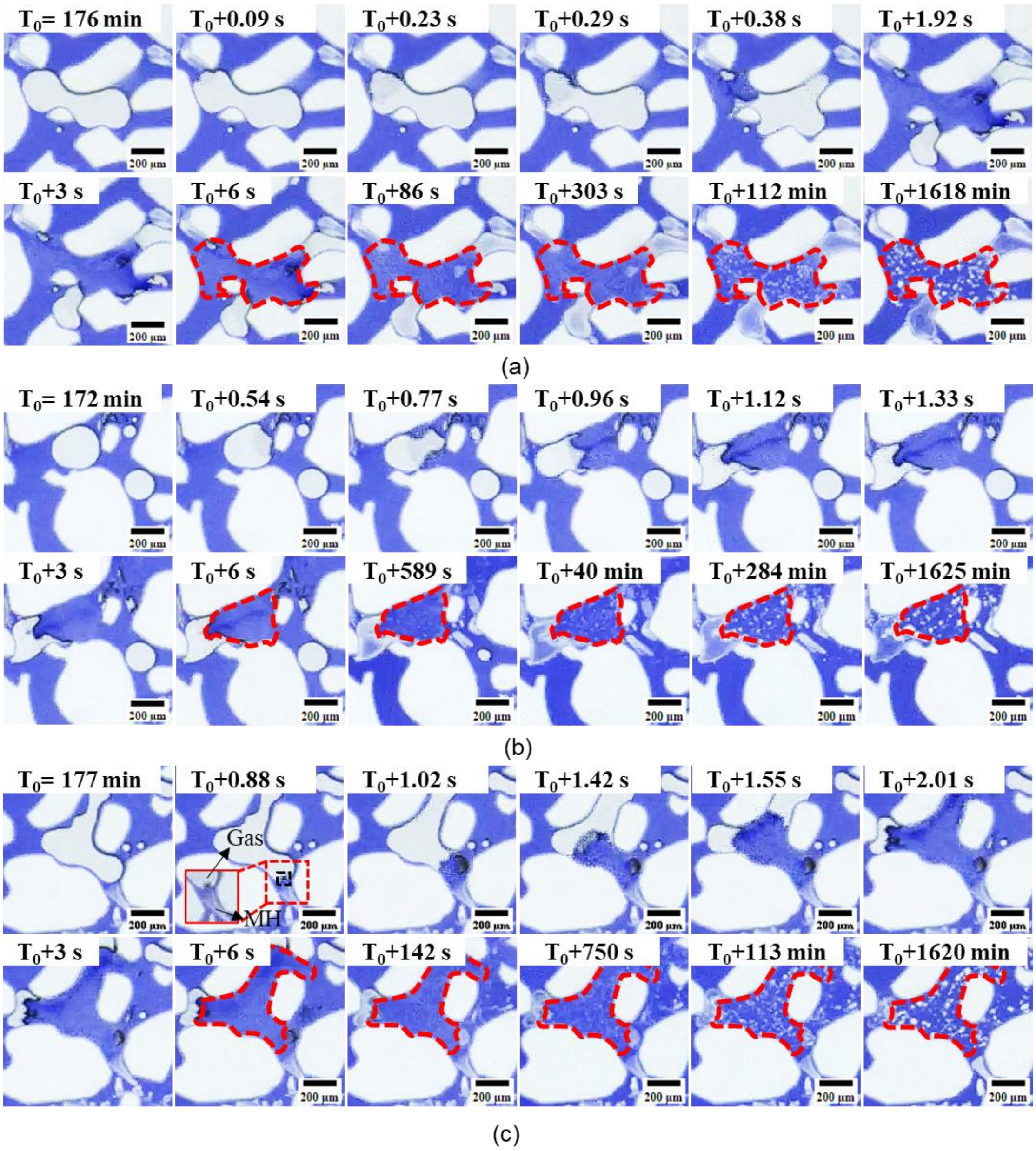

Figure 8 Formation procedure of point hydrates at three locations shown in (a-c) respectively

The quantification of the theorical and measured volume of point hydrates shown in Figure 8b is shown in Figure 9. The theorical volume of gas hydrate formed from the free gas phase was calculated based on decreased gas phase area, A1-A2, between before and after formation of point hydrate, A1 and A2, respectively. The volume of the consumed gas is calculated based on the area times the depth of the microfluidic chip. The number of moles of methane consumed can be calculated using the ideal gas law:

$$n_{CH_4} = \frac{P \cdot V_{CH_4}}{R \cdot T}$$

where P is the pressure (in Pa),$V_{CH_4}$is the volume of methane consumed (in m³), R is the gas constant (8.314 J/(mol K)), T is the temperature (in K).

The volume of gas hydrate can be calculated using the number of moles and the molar volume of methane hydrate:

$$V_{hydrate} = \frac{n_{hydrate} \cdot M_{hydrate}}{\rho_{hydrate}}$$

where$M_{hydrate}$ is the molar mass of methane hydrate (in g/mol), $\rho_{hydrate}$is the density of methane hydrate (typically around 0.9 g/cm$^3$).

The calculated area for point hydrate is via segmentation of image into black and white color first and then the hydrate volume is calculated based on the assumption that each of the point hydrate is hemisphere.

The total volume of point hydrates increases exponentially with time, and until about 1000 s the growth reaches the maximum value, which is about 80% of the theorical value. The difference between measured value and theorical value might be because some of the gas is dissolve into water.

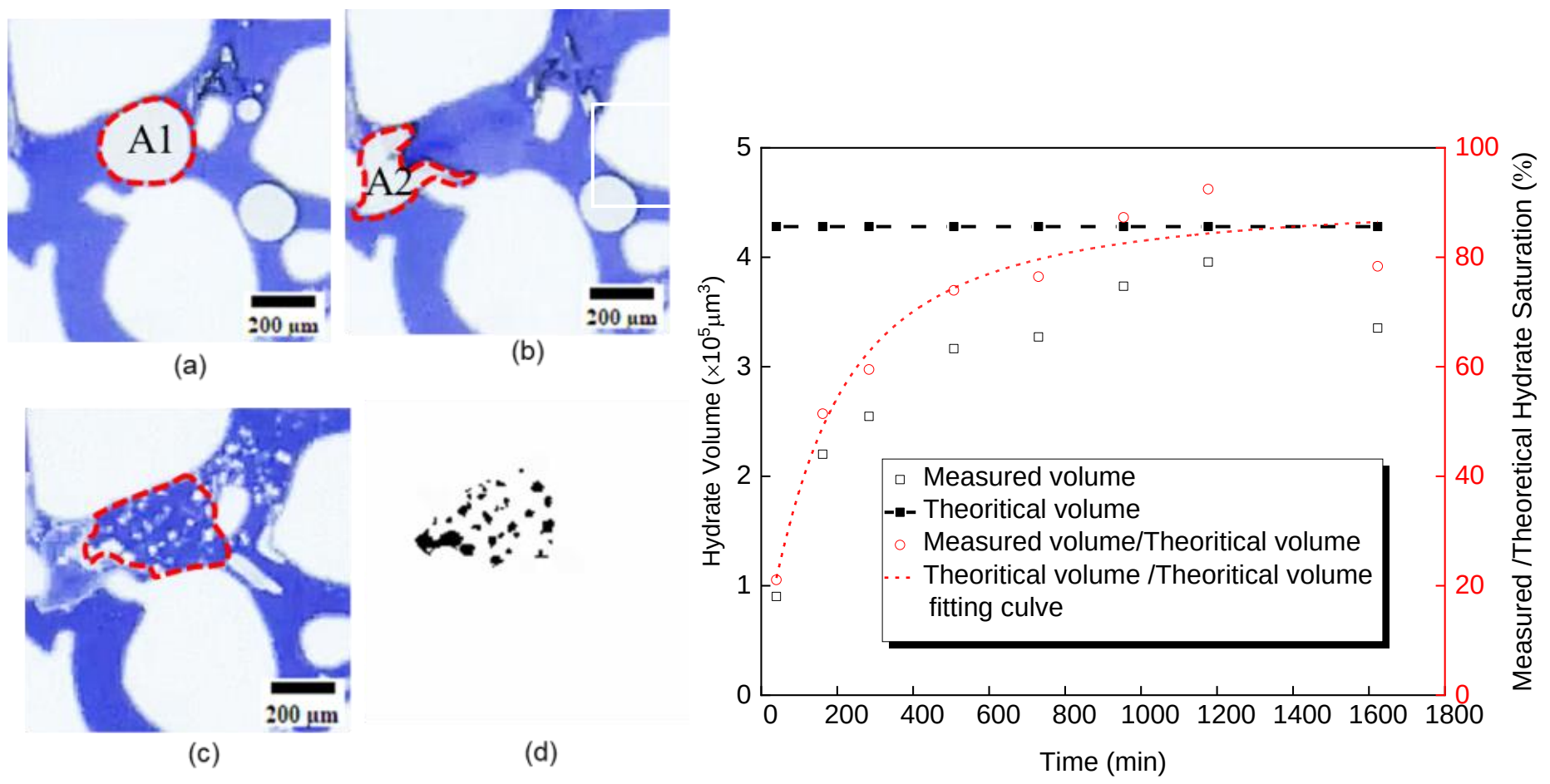

Figure 9 Quantification of the theoretical and measured volume of point hydrates shown in Figure 8b. The theoretical volume was calculated based on the decrease in gas phase area before and after the formation of point hydrate, while the measured hydrate volume was calculated assuming the hydrate is a hemisphere: (a) Initial free gas phase area A1; (b) Gas phase area after the formation of relocation A2; (c) Calculated area for point hydrate; (d) Segmentation of the image into black and white; (e) Hydrate volume and the ratio of measured to theoretical gas hydrate volume, $T_0$ = 173 min

**Formation of membrane hydrate in gas-water membrane at not-movable gas-phase**

Membrane hydrates form in the gas-water interface where the gas-phase methane remains immobile (Figure 10). Some free methane gas is trapped by adjacent pore structures, influenced by the Jamin effect and the hydrophilicity of the microfluidic chip surface. This creates a stable gas-liquid interface that effectively resists pressure fluctuations and deformation, making it difficult for the gas to move with the liquid phase (Figure 10). When the temperature and pressure conditions for hydrate formation are met, the water film between the gas and the microfluidic chip surface, which has a relatively high methane content due to constant exposure to free gas-phase methane, initiates hydrate formation. This process is similar to the formation of point hydrates. The key difference between the formation of membrane hydrates and point hydrates is that after the initial hydrate formation, the free gas-phase methane cannot migrate due to pore effects. As a result, dissolved methane in the water continuously transfers into the gap between the gas and the inner surface of the microfluidic chip or hydrates, both of which are hydrophilic (Zhang et al., 2024). Water migrates into the gap between the gas and the inner surface of the microfluidic chip or hydrates because the consumption of gas creates volume between the gas and hydrate, and the methane hydrate surface is hydrophilic. As water fills in, the color of the hydrate becomes bluer (Figure 10a, $T_0$+5~43 s; Figure 10b, $T_0$+12-36 s; Figure 10c, $T_0$+6~69 s). As the hydrates grow thicker, their color becomes less blue (Figure 10a, $T_0$+43~550 s; Figure 10b, $T_0$+36~342 s; Figure 10c, $T_0$+69~472 s). Eventually, the hydrates transition into a non-continuous form, similar to point hydrates but with much shorter distances between the pieces (Figure 10a, $T_0$+1634 min; Figure 10b, $T_0$+1624 min; Figure 10c, $T_0$+1635 min).

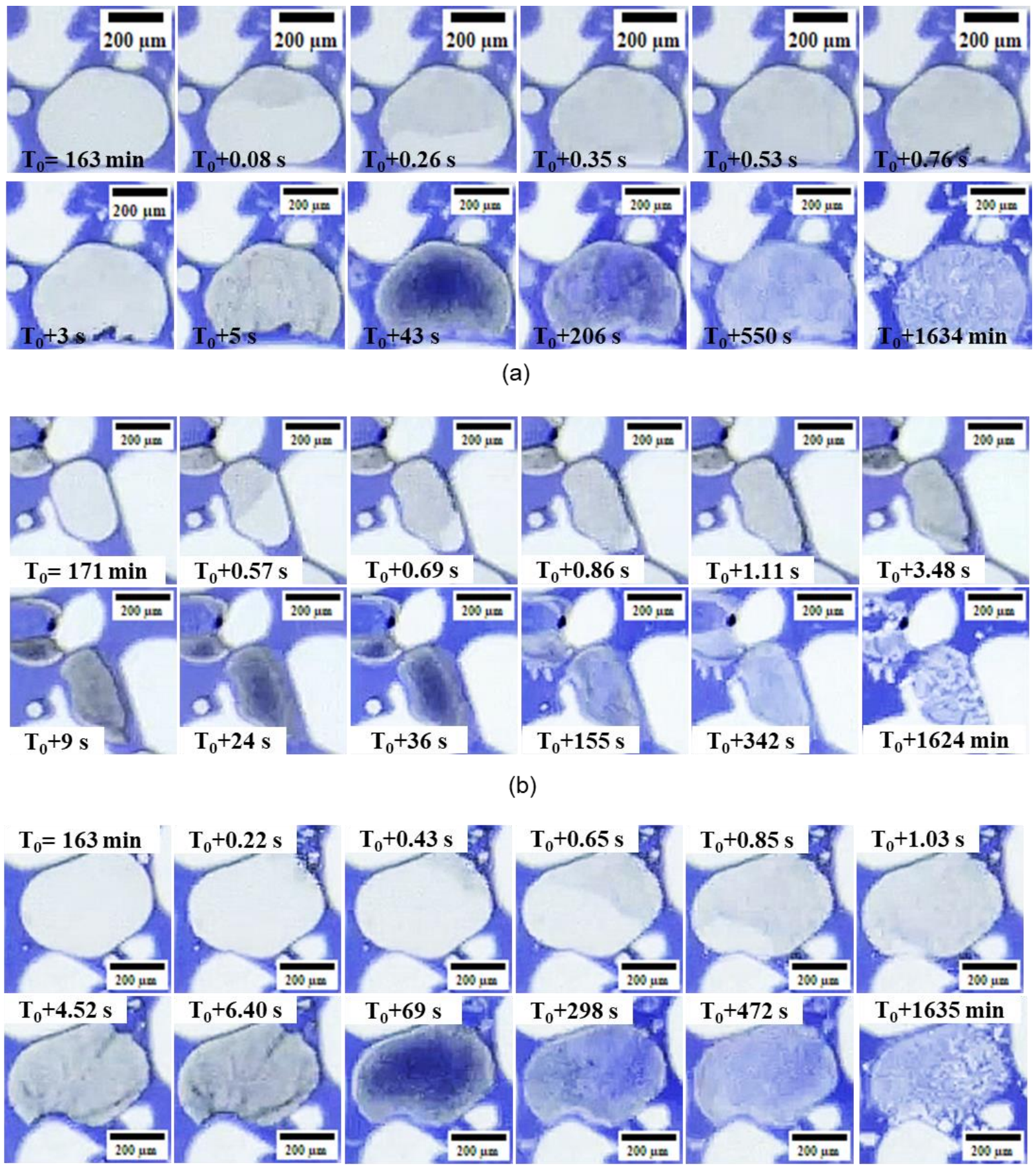

(c)

Figure 10 Formation procedure of membrane hydrates at three locations shown in (a-c) respectively

The quantification of membrane hydrate growth is illustrated in Figure 11. The original images (Figure 11a, top row) were processed to produce segmented images (Figure 11a, bottom row). In this segmentation, the blue regions were converted to white, and the non-colored areas were converted to black. The saturation of the crystals was then calculated using the pixel ratio of the black regions. Once the membrane hydrate starts growing, the growth procedure takes about 500s.

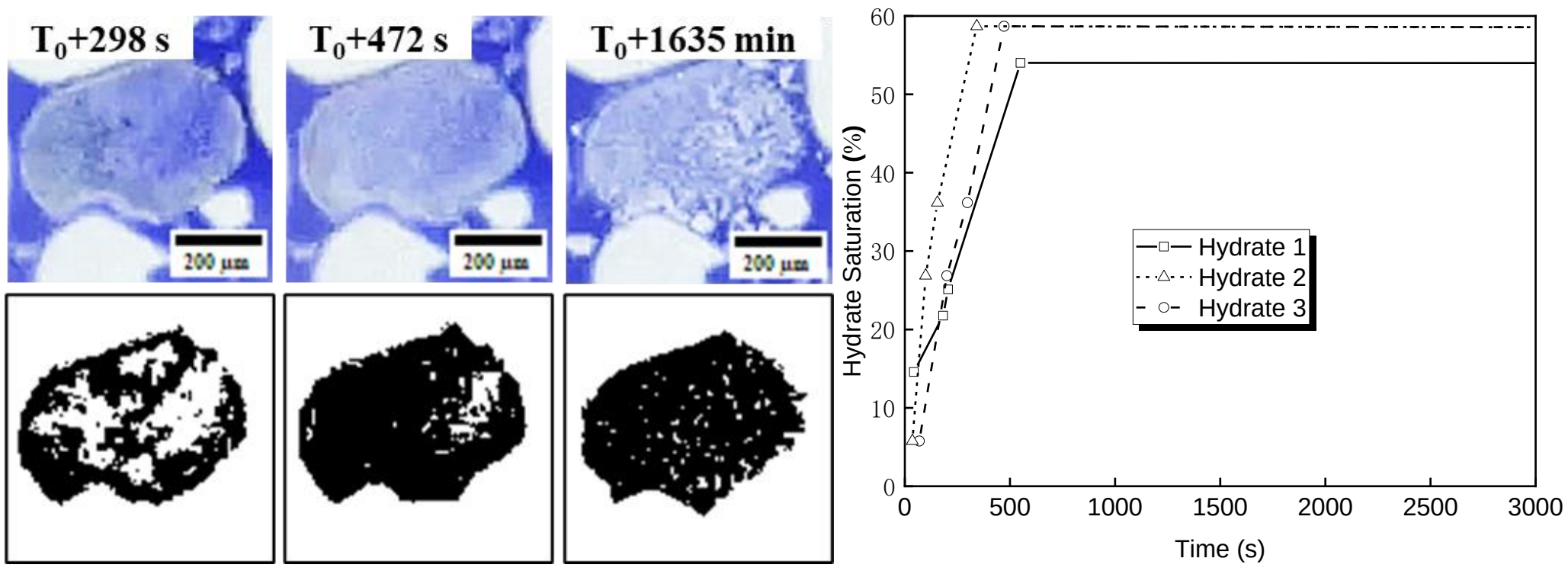

Figure 11 Quantification of membrane hydrates: (a) Images and segmented images of membrane hydrates at three time points; (b) Hydrate saturation (hydrate area relative to pore area) changes over time.

**Formation of shell hydrates surrounding migrated gas-phase**

Shell-like hydrates form after the migration of gas-phase methane into new pores (Figure 12a, $T_0$+0~6 s). As the gas migrates to a new location (Figure 12a, $T_0$+0~6 s), unlike in the formation of point and membrane hydrates, the concentration of dissolved methane in the water film between the gas and the microfluidic chip surface does not significantly exceed the concentration of methane in the imagined water membrane at the gas-water interface. Consequently, hydrates form uniformly around the gaseous methane under equally competitive conditions (Figure 12a, $T_0$+6 s~590 s), leading to a uniform hydrate shell around the bubble (Figure 12a, $T_0$+590 s~1674 min). This hydrate appears whiter around the bubble and light blue within it (Figure 12a). The white appearance is due to the thicker hydrate in the z direction (vertical), while the light blue color in the x-y direction (horizontal) indicates thinner hydrate layers with water phase saw through. The light blue color inside the hydrate suggests that water has penetrated it, indicating that the hydrate structure allows water to permeate. Because gas migration is highly dependent on pore shape, and the hydrate shell forms at the gas-water interface or within the water membrane, these crystals exhibit a pore-filling structure, albeit with water inside. Additionally, the shell hydrate is unstable and can dissolve when a more stable form of hydrate, such as block hydrate, forms (Figure 12b). This process can be explained by Ostwald ripening, which gradually alters the hydrate's pore habit from grain-attaching to pore-filling and increases hydrate saturation heterogeneity at both pore and core scales (Chen et al., 2018).

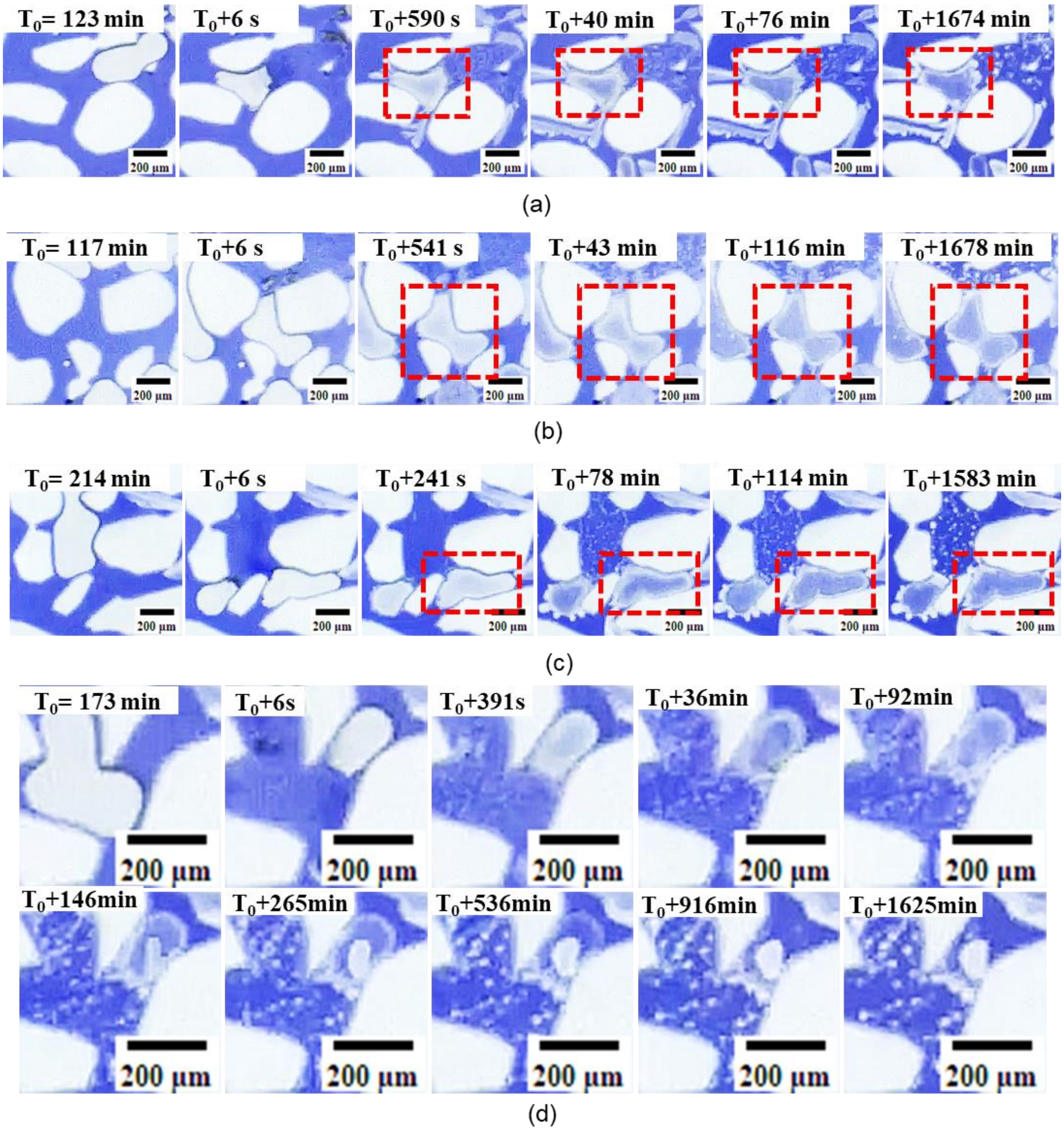

Figure 12 Formation procedure of shell hydrates at three locations shown in (a-c) respectively and instability of shell hydrates (d)

Previous studies have also suggested that methane bubbles form hydrate shells, which grow along the gas-water interface (Tohidi et al., 2001; Li et al., 2022a; Li et al., 2024) and extend toward the gas phase in a direction normal to the interface during hydrate formation (Ji et al., 2021; Li et al., 2022a; Li et al., 2022b; Li et al., 2024). Additionally, the morphology of these hydrates has been found to be discontinuous with cracked hydrate shells (Jung et al., 2012), porous with encapsulated methane gas (Almenningen et al., 2018), and composed of many tiny hydrate crystals (Li et al., 2024). The current study supports these findings, suggesting that pore-filling hydrates may originate from shell hydrates that form around migrated free gas and fill the pores of the porous medium.

**Mechanism of hydrate formation and transformation**

The formation procedure affected by pore structure and free gas methane distribution is shown in Figure 13. Both block and vein hydrates, crystal I and II, respectively, are generated by dissolved methane, growth in pore structure toward free gas methane (Figure 13a). Point hydrate, crystal III, is generated after the migration of gas phase, in water film between microfluidic chip and gas by water and the dissolved methane (Figure 13b). Shell hydrate, hydrate IV, is generated between the gas-water interface after the gas migrated to a new pore due to the production of point hydrate. During the formation of shell-like hydrate, water can penetrate the shell and continue forming hydrates. The shell-like hydrates are unstable, can be later change into blocky form of hydrates due to Ostwald ripping. Membrane hydrate, hydrate V, is generated by the dissolved methane in the membrane between microfluidic chip and gas, formed with the water invading gap between microfluidic chip and gas which stays in the pore. Membrane hydrate can eventually change into densely point-like crystals.

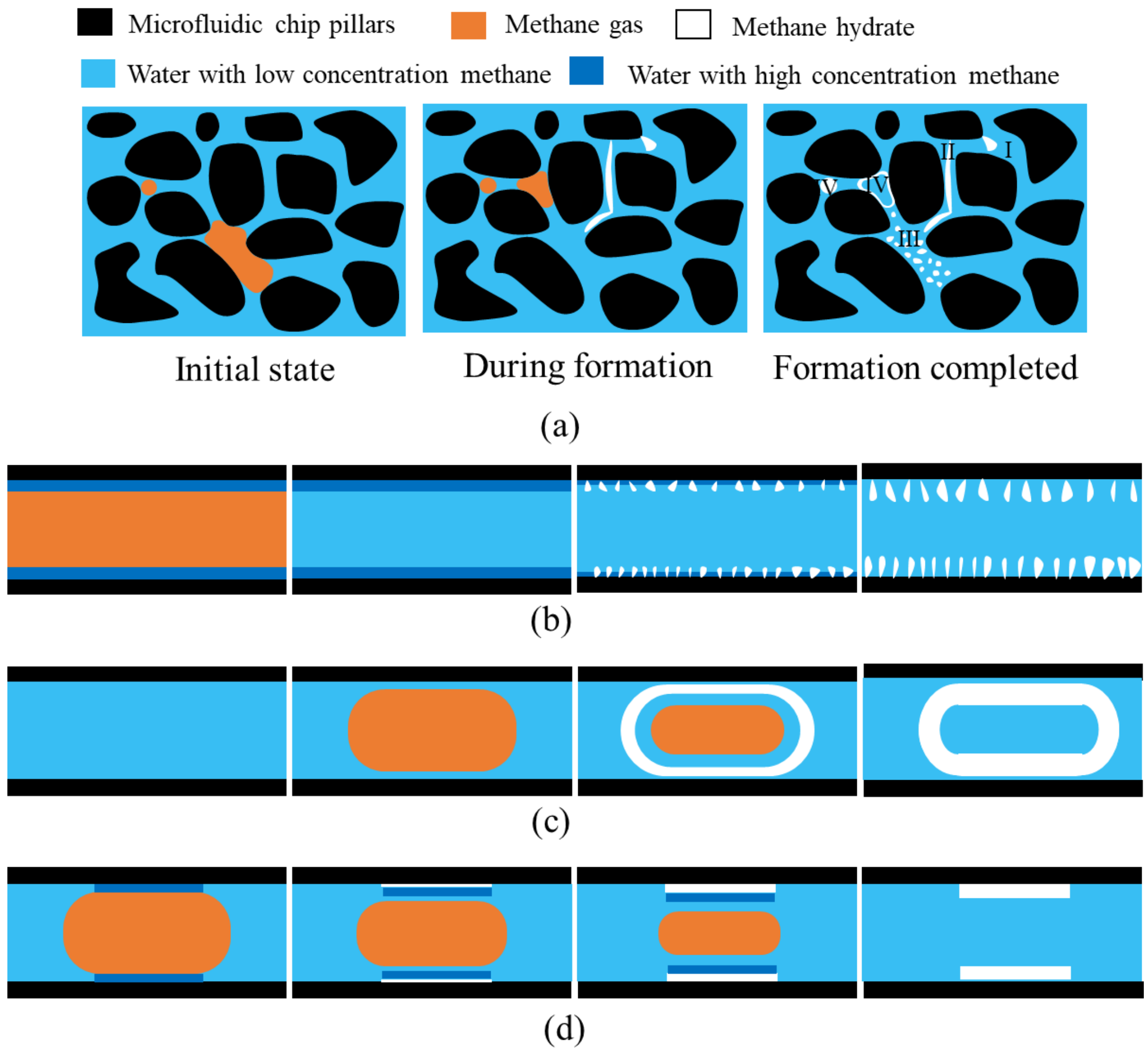

Figure 13 Schematic diagram of the formation of five types of hydrates: (a) Formation process of the five types of hydrates; (b) Formation process of point hydrate; (c) Formation process of shell hydrate; (d) Formation process of membrane hydrate.

The nucleation and growth procedure of hydrates have been studied but so far still being challenging. Studies found that the stochastic nature of nucleation makes the prediction of induction times an arduous endeavor and the nucleation time of methane hydrate, as an example, can range from 12 min to more than 24 h (Skovborg et al., 1993; Fandino et al., 2014). The growth can be described by three steps: 1) propagation of a hydrate thin film at the hydrocarbon/water interface, 2) film development, and 3) bulk conversion of the hydrate (Takeya et al., 2000; Taylor et al., 2007). The time range of each step is on the order of 10 s for propagation, 10 min for film development, and 10 h for bulk conversion. Chen et al. (2017) suggested that to complicate matters, hydrate crystallization, like many reactions occurring at gas–liquid–solid interfaces, is a complex process that can be influenced by intrinsic kinetics, mass transfer, and heat transfer. Ji et al. (2021) also found that methane hydrate formation does not occur at all bubbles at the same time but individually at different times indicating that the induction time of methane hydrate formation is different for different bubbles and hydrate formation in porous media is irregular. Following these studies, the current study found that the induction time of hydrates varied highly depend on hydrate types which are affected by the available form of methane and water. Membrane hydrates grow from water film between methane gas and microfluidic chip with constant quick supply of methane to dissolve into water film, therefore the formation time is quickest, which is only several seconds. Point hydrates grow from water film between methane gas and microfluidic chip after gas migrate away, without constant quick supply of methane to dissolve into water film, therefore, the formation time is slower than membrane hydrates, which is 40 to 112 mins. Shell hydrate grow from water film between the newly migrated methane gas and microfluidic chip or water, even though with constant supply of methane, but on one hand, the methane need time to dissolve into the water surround it, on the other hand, the formation of hydrate shell reduces the mass transfer between water and gas, the induction rate is therefore 40 to 78 mins.

Wang et al. (2021b) categorized hydrate morphologies based on gas-water ratios, identifying four key types. In excess-gas conditions, water forms thin films or resides at grain contacts, leading to grain-coating hydrates as gas diffuses into the water layer (Chen et al., 2018; Lei et al., 2019). At low gas-to-water ratios, gas forms dispersed bubbles in sediment pores, resulting in pore-filling hydrates (Tohidi et al., 2001). With moderate gas-to-water ratios, hydrates grow at the gas-water interface, leading to mixed grain-coating and pore-filling morphologies (Chaouachi et al., 2015). Under dissolved-gas conditions, hydrates nucleate in pore centers and eventually attach to grain surfaces, forming a combined morphology (Tohidi et al., 2001; Waite et al., 2013; Kuang et al., 2020). Building on these studies, we found that block and vein hydrates can exhibit either grain-coating or pore-filling morphologies, while shell hydrates initially exhibit pore-filling morphology before transitioning to vein hydrates. Point hydrates and grain coatings are primarily grain-coating. Kang et al. (2016) predicted that grain-coating hydrates exhibit higher water permeability than pore-filling hydrates at the same saturation. Future research should focus on correlating the factors influencing the distribution of these hydrate types and their impact on sediment permeability.

**Hydrate decomposition process and effects of gas phases**

When the pressure drops to the hydrate decomposition pressure, fine bubbles first form at the edges of the hydrate (Figure 14 $T_0$=560 s). The gas bubble keeps growing (Figure 14 $T_0$+10 s~ $T_0$+40 s) and could potentially merged with other free gas phases (Figure 14 $T_0$+45 s). During the desiccation procedure, the gas and water phases keep moving due to the local pressure change in the microfluidic chip (Figure 14). Despite the changes of gas water phases, the general obvious dissociation of gas hydrate happens when it is attached to gas phases (Figure 14). Without the gas phases touching hydrates, hydrate dissociation rate is low (Figure 14 $T_0$+90 s~ $T_0$+110 s, gas on the right side of the hydrate crystal).

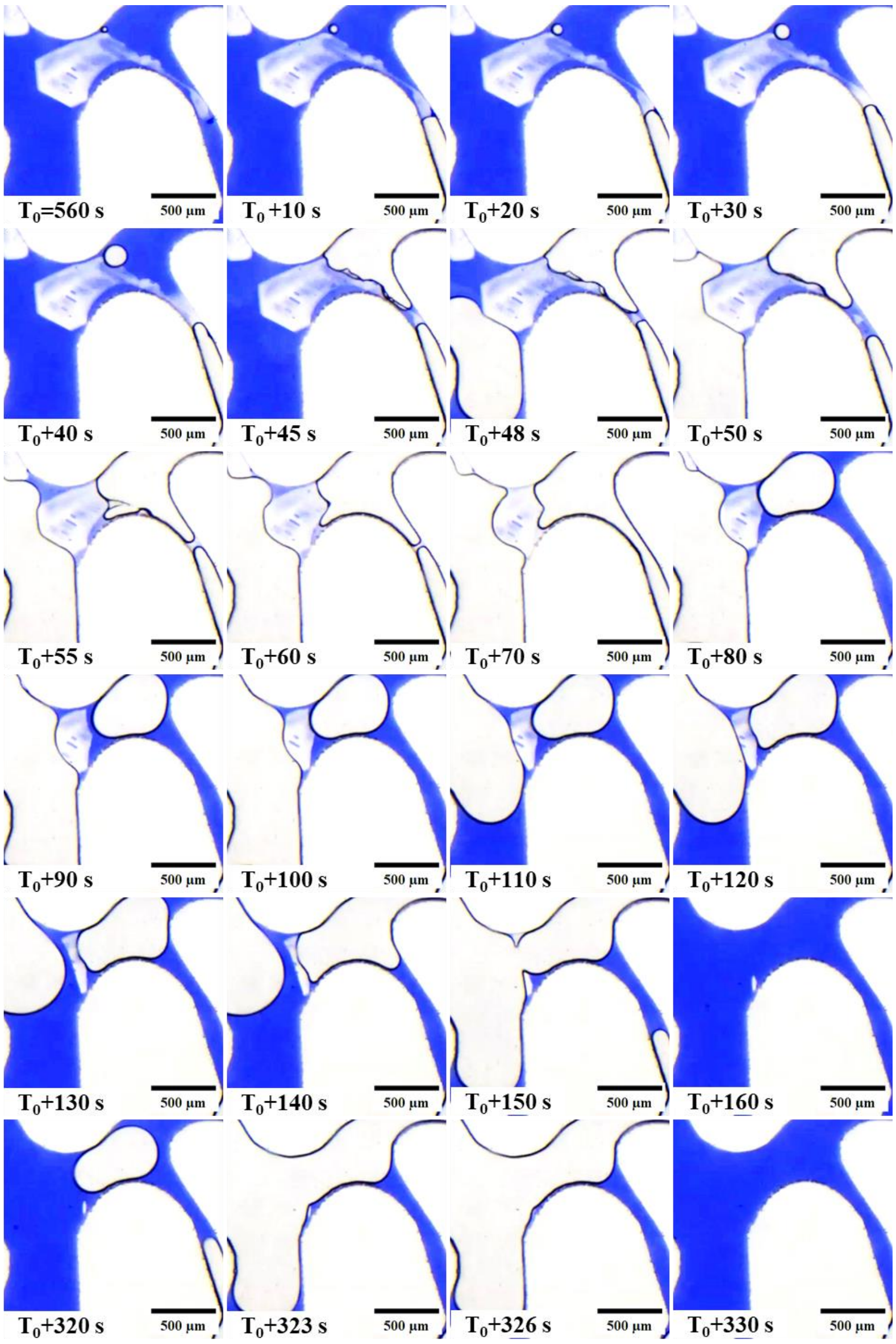

Figure 14 Dissociation process of block hydrate

During depressurization-induced decomposition, the hydrate decomposition rate in contact with the gas phase is significantly higher than that in contact with the liquid phase. The effects of water and gas on hydrate decomposition rates are analyzed based on Figure 15a, and the results are shown in Figure 15 b and c. When hydrates are in contact with the gas phase and liquid phase, respectively, the decomposition rates are significantly different (Figure 15 b and c). During $T_0$=11 min~$T_0$+92 s, $T_0$+237 s~$T_0$+323 s and $T_0$+1606 s ~ $T_0$+1810 s, there are gas in contact with the hydrates, and the area of hydrate reduce with time significantly faster than when there is only water in contact with water (Figure 15b, $T_0$+92 s~$T_0$+237 s, $T_0$+323 s~ $T_0$+1606 s). The decomposition rate of hydrates in contact with the gas phase is approximately 12 times that of hydrates in contact with the liquid phase (Figure 15 c). Further quantitative experimental results (Figure 15d) indicate the areas of the three hydrates in the observation area decrease with time at different rates (Figure 15e), there is a liner correlation between the hydrate decomposition rate and the gas- hydrate contact area (Figure 15f).

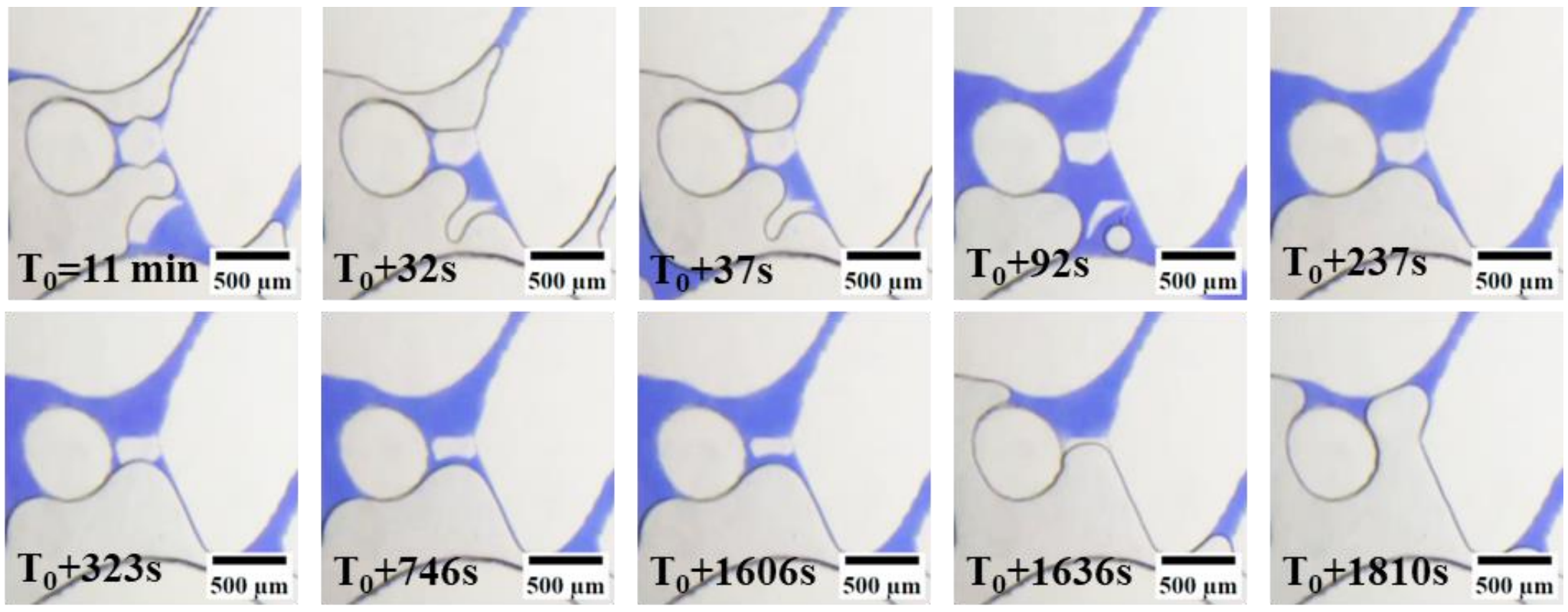

(a)

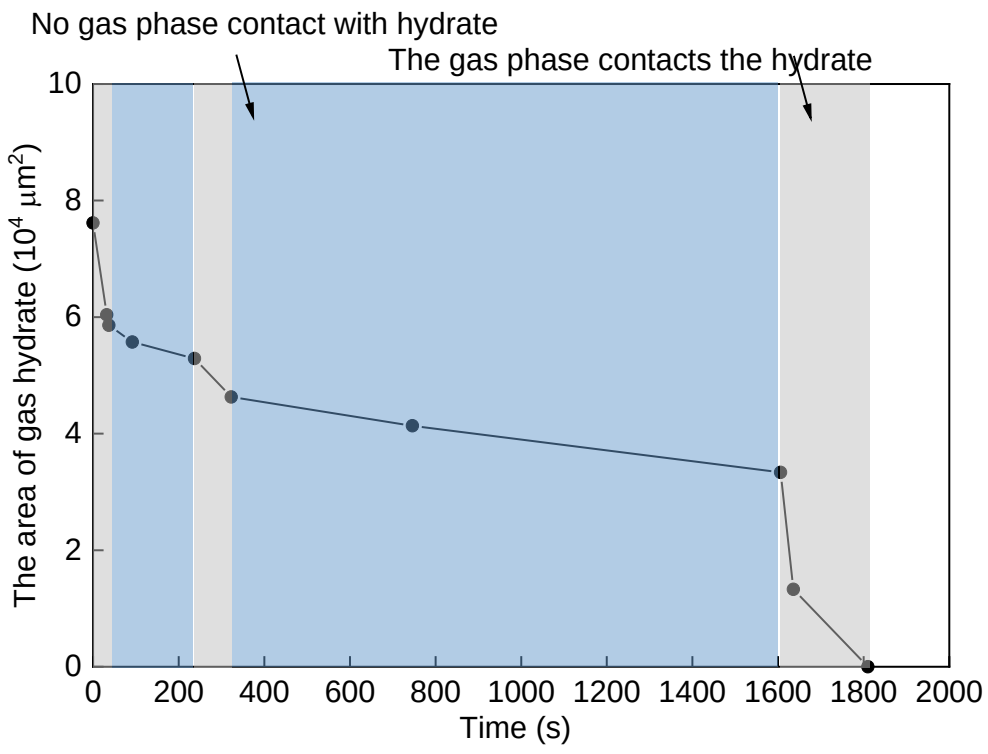

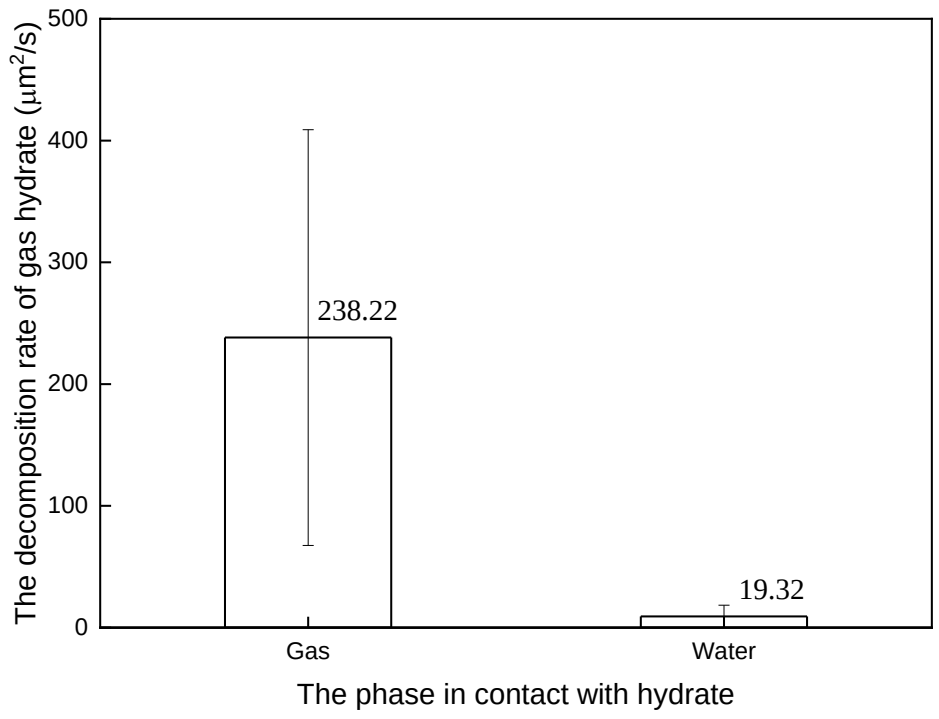

(b) (c)

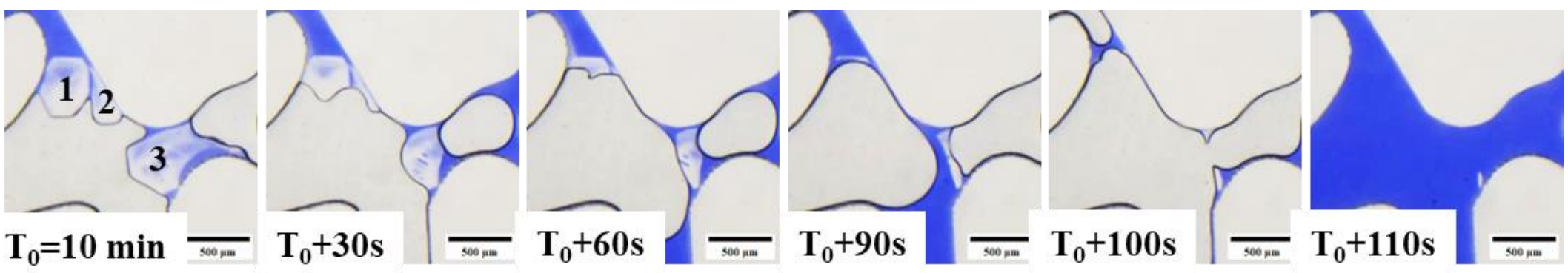

(d)

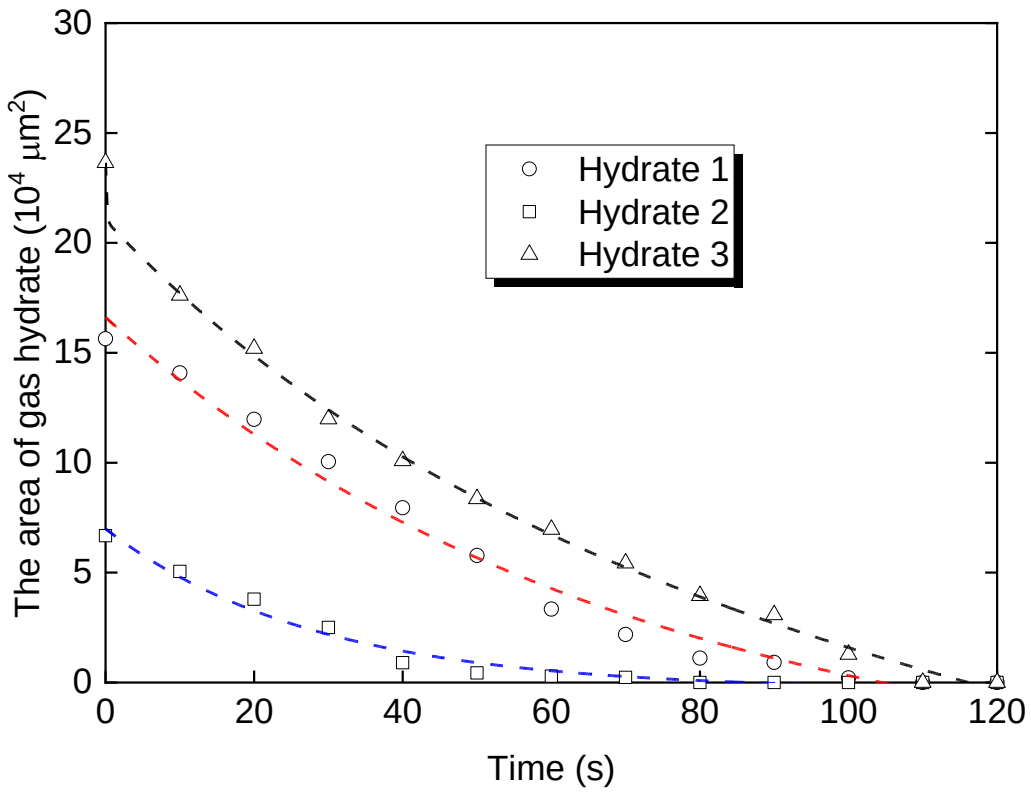

(e)

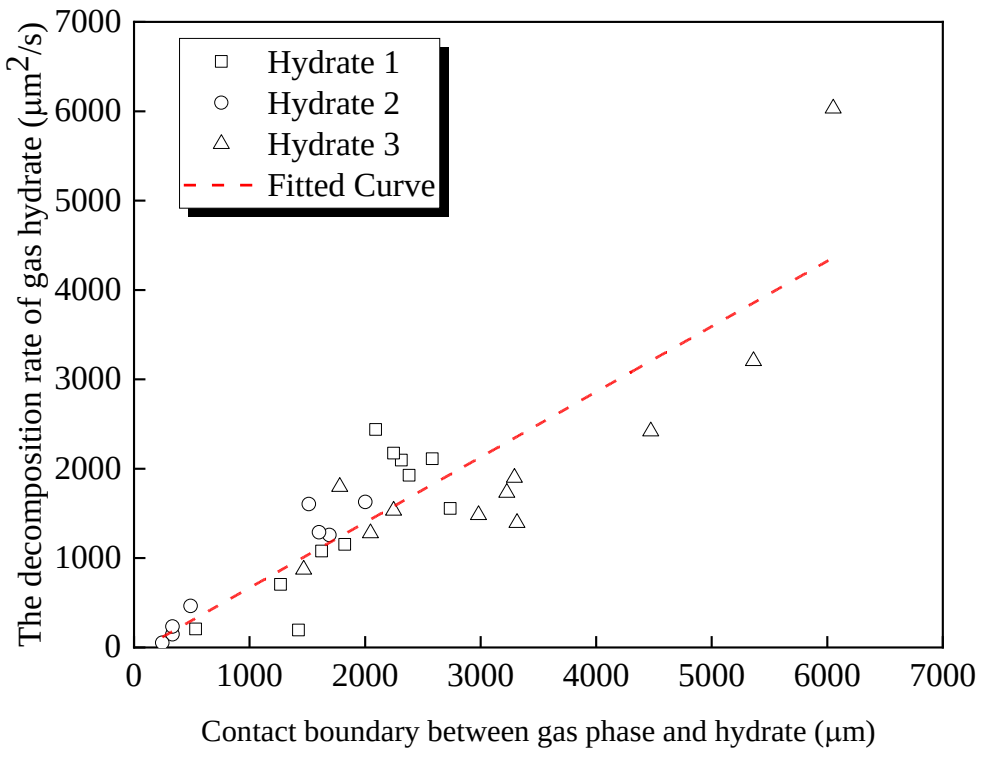

(f)

Figure15 Effects of water and gas on hydrate decomposition rates: (a) Dissociation process of block hydrate under the influence of water and gas; (b) Hydrate area decrease over time; (c) Decomposition rate of gas and hydrate when in contact with gas and water; (d) Decomposition process of hydrate in contact with gas; (e) Hydrate area decrease over time; (f) Correlation between hydrate decomposition rate and contact length between gas and hydrate.

Yang et al. (2024) suggested that while excess dissolved methane and released methane from melted hydrate promote gas bubble growth, these bubbles significantly accelerate hydrate dissociation, with spontaneous gas-water migration playing a minor role in the dissociation rate. Consistent with this study, our research also found that the gas phase accelerates the dissociation of hydrate. Additionally, our study revealed that the effect of gas-phase contact with hydrate is approximately 12 times greater than that of water-phase contact. The dissociation rate of gas hydrate is linearly correlated with the contact boundary area between gas and gas hydrates.

Hydrate dissociation occurs either in the presence of a contacting gas phase or through the formation of a gas bubble (Figures 15 and 17b). As the hydrate dissociates, the gas phase expands. When the gas is removed from the vicinity of the hydrate, dissociation nearly ceases (Figures 16a and 17b, $T_0$+200s). The rate of gas generation from block hydrates is faster than that from point hydrates. This difference may be attributed to the fact that point hydrates are not interconnected, requiring the gas to diffuse through water before merging into a continuous gas phase. In contrast, gas generated from block hydrates merges directly into the gas phase, resulting in a quicker release.

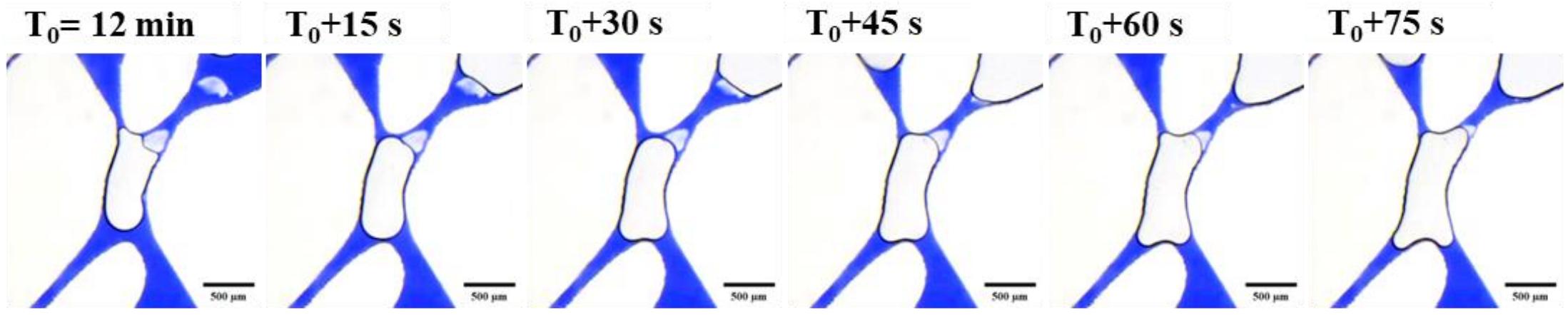

(a)

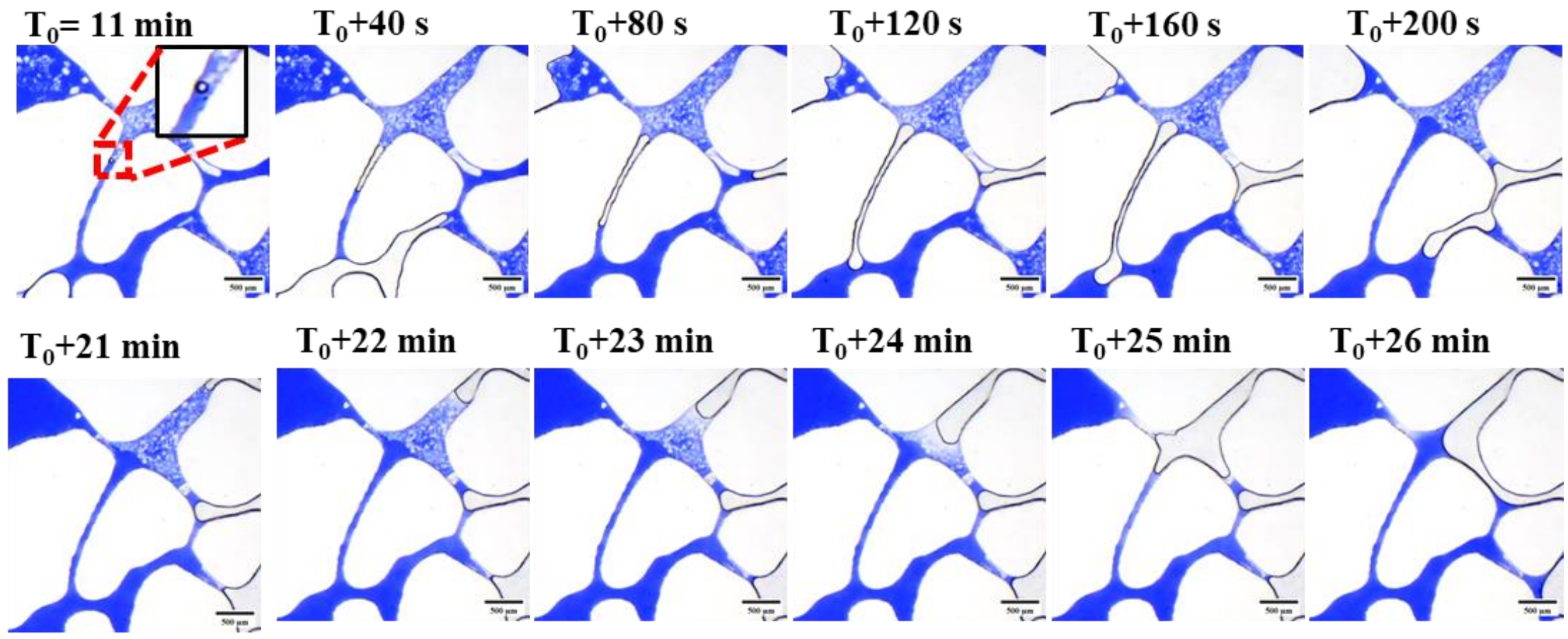

(b)

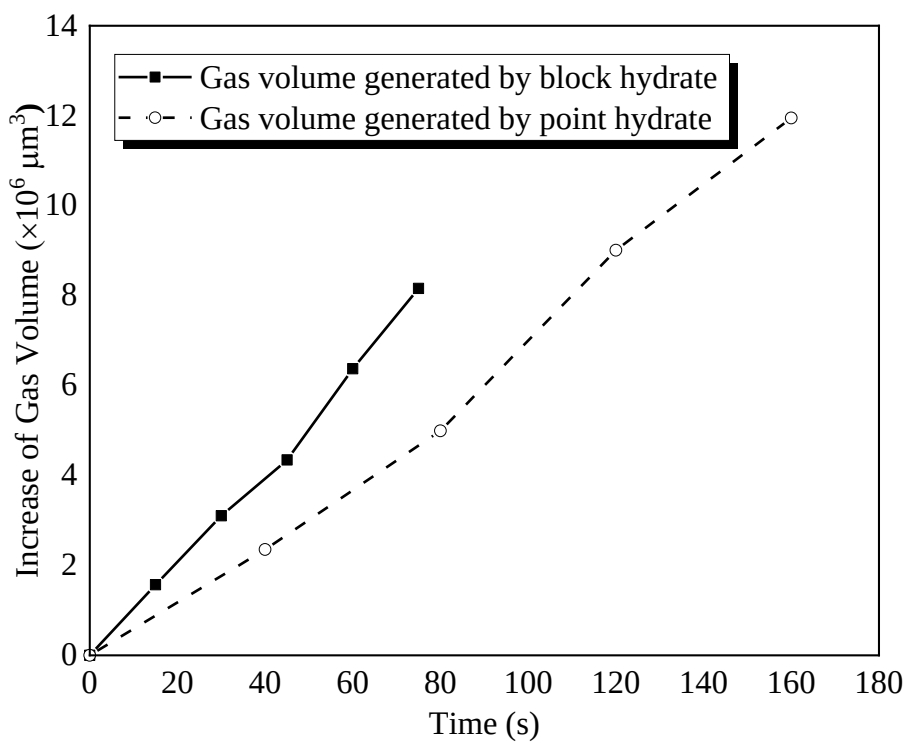

(c)

Figure 16 Dissociation process of block hydrate and point hydrate: (a) Dissociation process of block hydrate and point hydrate; (b) Dissociation process of block hydrate and point hydrate; (c) Increase in gas volume generated from block hydrate and point hydrate over time.

**The migration of liquid, gas, and solid during the decomposition of hydrates**

The initial liquid-gas distribution determines the distribution of different characteristic hydrates. The locations where gas is present, compared to those where liquid is present, are non-primary flow channels, leading to the formation of spot-like hydrates in these areas, while block-like hydrates form in the primary channels. When hydrate dissociation occurs, the block-like hydrates in the primary channels preferentially dissociate under the influence of pressure differences. The resulting gas phase tends to migrate first to the primary flow channels, further promoting the preferential dissociation of block-like hydrates. The gas then flows into the secondary channels, where spot-like hydrates had previously formed in greater quantities. Under the combined effects of pressure differences and the gas phase, these spot-like hydrates dissociate. Therefore, the overall dissociation of spot-like hydrates occurs later than that of block-like hydrates (Figure 17a). The water produced by hydrate decomposition moves towards the liquid phase, while the gas produced moves towards the gas phase (Figure 17 b). The disturbance caused by depressurization can lead to the migration of some small particles of natural gas hydrates (Figure 17 c).

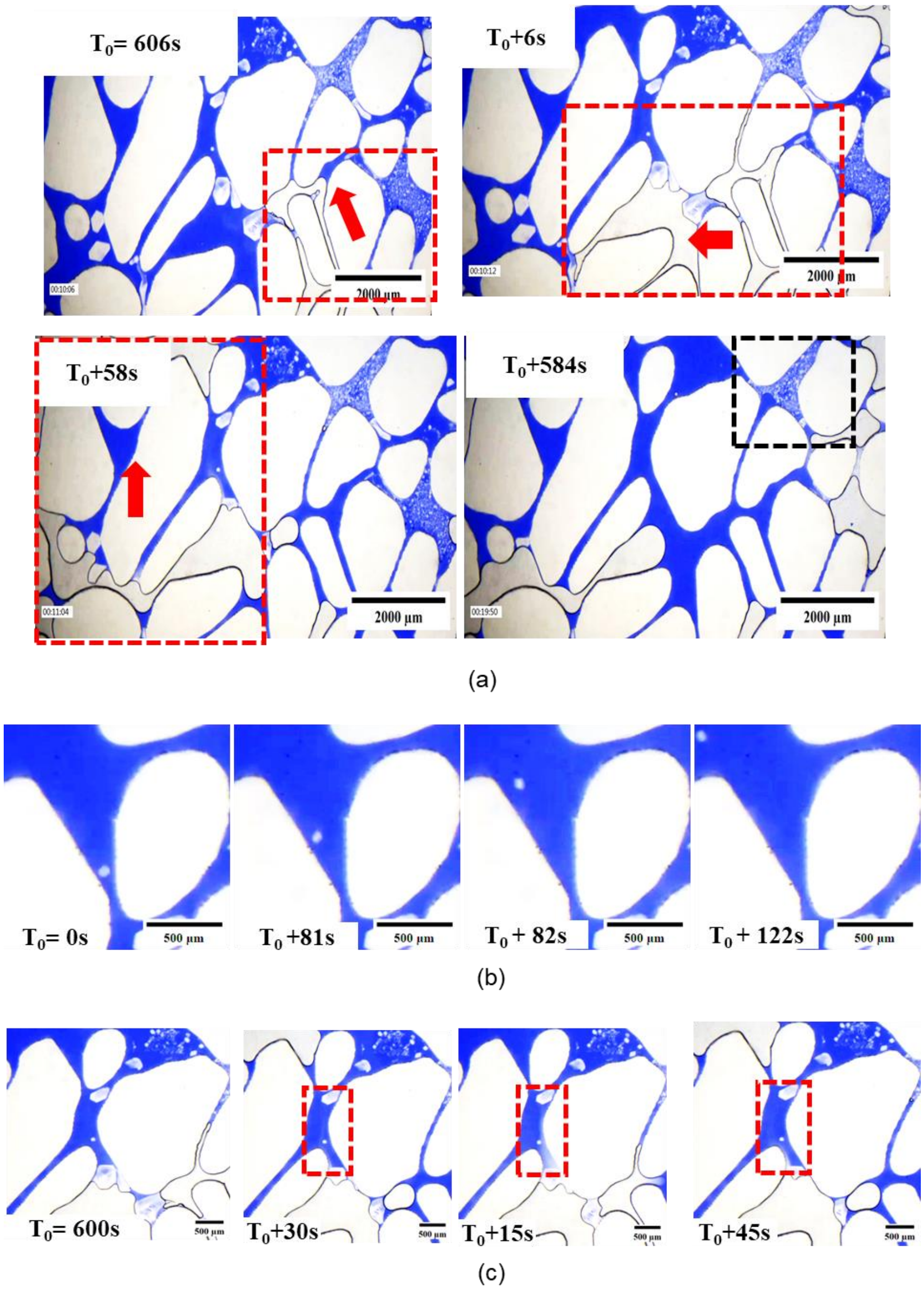

Figure 17 Hydrate depressurization decomposition process: (a) The primary seepage channel decomposes first (red box), while the secondary seepage channel decomposes last (black box); (b) Small hydrate particles migrate during the initial stage of depressurization; (c) The flow pattern of gas and liquid during depressurization decomposition.

Previous studies have illustrated that during the dissociation process, the average diameter of the bubbles increased with time, and the total number of bubbles decreased with time (Wang et al., 2021b). In addition, three distinct stages of gas bubble evolution were identified during MH dissociation via thermal stimulation: (a) single gas bubble growth with an expanding water layer at an initial slow dissociation rate, (b) rapid generation of clusters of gas bubbles at a fast dissociation rate, and (c) gas bubble coalescence with uniform distribution in the pore space (Zhang et al., 2024). Following these studies, our research found that during dissociation, gas hydrates first dissociate in the primary flow channels, leading to the formation of gas. This gas then migrates through the primary flow channels, where block hydrates are mainly distributed, to the secondary channels, where point hydrates predominantly occur.

# 4. Conclusions

In this study, microfluidic technology, which effectively simulates natural conditions and sedimentary porous media, together with the use of methylene blue, which enhance phase differentiation, was applied to study the methane hydrate formation and decomposition processes, characteristics and kinetics.

Five distinct types of methane hydrates were identified: block, vein, point, membrane, and shell hydrates. The distribution and movement of free gas phases were found to play a crucial role in shaping these hydrate structures, influencing their morphology, direction, and stability. The research demonstrated that block and vein hydrates primarily form within water phases, driven by dissolved gas, and typically manifest as pore-filling or bonding types. In contrast, point and membrane hydrates are closely linked to the dynamics of free gas phases, forming as coatings on microfluidic surfaces, akin to their formation on sediment particles in natural environments. Shell hydrates, which develop as pore-filling structures after gas migration, further illustrate the complexity and variability of hydrate morphology in response to gas movement.

The findings underscore the significant role of free gas phases in both the formation and dissociation of methane hydrates. Free gas phases dictate the direction of block and vein hydrate growth, with a strong correlation between the amount of free gas and the extent of hydrate formation. The movability of the free gas phase also impacts the formation of point, membrane, and shell hydrates. Specifically, membrane hydrates form when the free gas phase is immobile, while point and shell hydrates form when gas migrates, indicating a strong relationship between gas phase dynamics and hydrate morphology.

The dissociation process of hydrates was found to be strongly influenced by gas bubble formation. The movement of gas and water phases significantly accelerates the dissociation rate—up to approximately 12 times faster compared to dissociation in the presence of water alone. Block hydrates located in primary flow channels dissociate more rapidly than point hydrates in non-primary channels, highlighting the importance of hydrate location and the dynamic behavior of gas phases in determining dissociation kinetics. Moreover, the migration of gas and water phases during dissociation, along with the potential displacement of small hydrate fragments by pressure gradients, adds further complexity to the decomposition process.

During hydrate dissociation, gas bubbles are generated at both block and point hydrates, leading to accelerated dissociation as the gas phase expands. The continual movement of gas and water phases during dissociation further influences the process, with hydrates continuing to dissociate when in contact with the gas phase but significantly slowing when the gas phase moves away. The study also observed that gas phases tend to migrate toward other gas phases, and water phases toward water, during dissociation. Additionally, small hydrate fragments generated during dissociation may migrate through the porous medium due to pressure gradients, potentially impacting the surrounding environment.

Overall, this research provides a comprehensive understanding of the mechanisms governing methane hydrate formation and dissociation in sedimentary porous media. The insights gained here are critical for optimizing natural gas extraction processes and mitigating the risks of pipeline blockages in deep-sea environments, offering a valuable foundation for future studies in this area.